\documentclass[a4paper, twocolumn, accepted=2021-08-19]{quantumarticle}
\pdfoutput=1

\usepackage{graphicx}
\usepackage{multirow}
\usepackage{amsmath}
\usepackage{amsfonts}
\usepackage{amsthm}
\usepackage{soul}
\usepackage{verbatim}
\usepackage{bm}
\usepackage{outlines}
\usepackage{siunitx}
\usepackage[space]{grffile}
\usepackage[colorlinks=true, linkcolor=blue, citecolor=gray]{hyperref}
\usepackage[capitalize]{cleveref}
\usepackage{xr}
\usepackage{color}
\usepackage{tikz}
\usepackage{braket}
\usepackage{enumitem}
\usepackage{cancel}
\usepackage{float}
\usetikzlibrary{calc, positioning}

\newcommand{\<}{\langle}
\renewcommand{\>}{\rangle}
\newcommand{\chat}{\hat{c}}
\newcommand{\chatdag}{\hat{c}^{\dag}}

\newcommand{\II}{\mathbb{I}}
\newcommand{\Hh}{\mathcal{H}}

\newcommand{\NN}{\mathbb{N}}
\newcommand{\RR}{\mathbb{R}}

\newcommand{\PP}{\mathbb{P}}
\newcommand{\Ss}{\mathcal{S}}
\newcommand{\Ll}{\mathcal{L}}
\newcommand{\Kk}{\mathcal{K}}
\newcommand{\Mm}{\mathcal{M}}
\newcommand{\Nn}{\mathcal{N}}

\newtheorem{thm}{Theorem}
\newtheorem{dfn}[thm]{Definition}
\newtheorem{exm}[thm]{Example}
\newtheorem{lem}[thm]{Lemma}

\newcommand{\numqubits}{{N_q}}
\newcommand{\numparams}{{N_p}}
\newcommand{\numunitaries}{{N_u}}
\newcommand{\numcouplings}{{N_c}}

\begin{document}

	\title{A diagrammatic approach to variational quantum ansatz construction}
	\author{Y. Herasymenko}
	\author{T.E. O'Brien}
	\affiliation{Instituut-Lorentz, Universiteit Leiden, P.O. Box 9506, 2300 RA Leiden, The Netherlands}
	
	\begin{abstract}
		Variational quantum eigensolvers (VQEs) are a promising class of quantum algorithms for preparing approximate ground states in near-term quantum devices.
		Minimizing the error in such an approximation requires designing ansatzes using physical considerations that target the studied system.
		Coupled-cluster ansatzes family are one family of such ansatzes, that are based on the perturbative principle of size-extensivity. Informally this principle means that the ground state quantum correlations are to be compactly represented in the ansatz.
		Unfortunately the size-extensive ansatzes usually require expansion via Trotter-Suzuki methods to be used on digital quantum computers. These introduce additional costs and errors to the variational approximation.
		In this work, we present a diagrammatic scheme for digital VQE ansatzes, which is size-extensive but does not require Trotterization.
		We start by designing a family of digital ansatzes that explore the entire Hilbert space with the minimum number of free parameters.
		We then demonstrate how one may compress an arbitrary digital ansatz, by enforcing symmetry constraints of the target system, or by using them as parent ansatzes for a hierarchy of increasingly long but increasingly accurate sub-ansatzes.
		We apply a perturbative analysis and develop a diagrammatic formalism 
		that ensures the size-extensivity of generated hierarchies.
		We test our methods on a short spin chain, finding good convergence to the ground state in the paramagnetic and the ferromagnetic phase of the transverse-field Ising model.
	\end{abstract}
	\maketitle

	\section{Introduction}
	Despite promises of exponential speedups, quantum algorithms require optimization to achieve an advantage over their classical counterparts on state of the art supercomputers for problems of interest.
	This is the case both in the Noisy Intermediate-Scale Quantum era~\cite{Preskill18}, where coherence times in quantum devices prohibit all but the shortest experiments to be performed, and in first-generation fault-tolerant devices, where a single non-Clifford rotation requires thousands of additional qubits and hundreds of error correcting cycles~\cite{Litinski19}.
	In the field of digital quantum simulation, the variational quantum eigensolver (VQE)~\cite{Peruzzo14} has emerged as a competitive class of algorithms for generating approximate ground states of quantum systems, due to its relatively low circuit length.
	These algorithms consist of parametrizing a quantum circuit with a small number of classical control variables, which may be tuned to minimize the energy of the state produced by the circuit, given a target Hamiltonian.
	As the manifold of obtainable states for a given VQE will only ever be an exponentially small region in the larger Hilbert space, optimizing VQE design is critical to obtain good approximations of the system's ground state~\cite{Mcclean16,Romero18,McClean18}.
	This has spurred much recent work in optimizing VQEs based on the unitary coupled cluster expansion~\cite{Mcclean16,Romero18,DallaireDemers18}, or on the quantum approximate optimization algorithm~\cite{Farhi14,Lloyd18}.
	The efficiency of coupled cluster methods is based on the principle of size-extensivity. This means that the ansatz systematically accounts for ground state correlations, as ensured in perturbative language by the linked-cluster theorem \cite{brueckner55}.
	However, to be realized as a quantum circuit size-extensive ansatzes typically require expansion via Trotter-Suzuki-based methods~\cite{trotter1959product,suzuki1991general}. At low circuit depth, these expansions introduce significant errors.
	Alleviating this issue would help to ensure the efficiency of the VQE algorithm.	
	
	In this work, we develop a Trotterization-free diagrammatic method to generate size-extensive VQEs. 
	We start by designing a class of VQE ansatzes, based on the stabilizer formalism in quantum error correction, which provably tightly span the entire Hilbert space of $\numqubits$ qubits.
	We then demonstrate how one may compress an arbitrary variational ansatz to account for symmetries of a target Hamiltonian.
	We further show how to construct a hierarchy of ansatz generators, allowing one to trade between circuit length and accuracy in a practical manner by choosing only those generators that contribute well to solving the problem.
	We motivate the construction of one particular such hierarchy from a general perturbative analysis of weakly coupled target Hamiltonians, for which we develop a simple-to-use diagrammatic formalism. 
	We find that our geometrically tight stabilizer ansatz may be compressed to a practical size using this perturbative scheme. 
	The analogue of the linked-cluster theorem for such compressed digital ansatzes is stated and proven, ensuring the size-extensivity of the construction.
	We also propose some possible modifications to our perturbative scheme to account for circuit depth and locality.
	We compare the performance of these constructions on simulations of the transverse-field Ising model in three different physical regimes (weak-coupling, strong-coupling, and critical).
	We find that strictly following the perturbative approach is beneficial in the weak-coupling regime, but restricting the ansatz to lowest-order gives better convergence in the strong-coupling regime --- even though such ansatzes are seemingly less-informed about the strong-coupling physics.
	
	\section{Variational quantum eigensolvers}
	
	A variational quantum eigensolver (VQE) is an algorithm executed on a quantum register that aims to approximate the minimum eigenvalue $E_0$ of a target Hamiltonian $H$ on $\mathbb{C}^{2^{\numqubits}}$ by finding low energy states $|\psi\>\in\mathbb{C}^{2^{\numqubits}}$ variationally.
	To be precise, this algorithm minimizes $\<\psi|H|\psi\>$ over a variational ansatz:
	
	\begin{dfn}\label{dfn:var_ansatz}
		A variational ansatz on $\numparams$ parameters corresponds to a pair $(U,|\vec{0}\>)$, where $U$ is a smooth map from the \textbf{parameter space} $\vec{\theta}\in\mathbb{R}^{\numparams}$ to the unitary operator $U(\vec{\theta})$ on $\mathbb{C}^{2^{\numqubits}}$, and $|\vec{0}\>\in\mathbb{C}^{2^{\numqubits}}$ is the \textbf{starting state}, which is acted on to generate the \textbf{variational state} $|\psi(\vec{\theta)}\>=U(\vec{\theta})|\vec{0}\>$, with variational energy $E(\vec{\theta})=\<\psi(\vec{\theta})|H|\psi(\vec{\theta})\>$.
	\end{dfn}

	As a brief example, let us define the following toy two-qubit variational ansatz:

	\begin{exm}
		The $3$-parameter \textbf{YYX} variational ansatz $(U_{YYX},|00\>)$ is defined on two qubits $\{Q_1,Q_2\}$, with the starting state $|00\>$ in the computational ($Z$) basis, and
		\begin{equation}
			U_{YYX}(\theta_1,\theta_2,\theta_3):=e^{i\theta_3Y_1X_2}e^{i\theta_2Y_2}e^{i\theta_1Y_1}.\label{eq:UYYX_def}
		\end{equation}
	\end{exm}

\begin{figure}
	\includegraphics[width=\columnwidth]{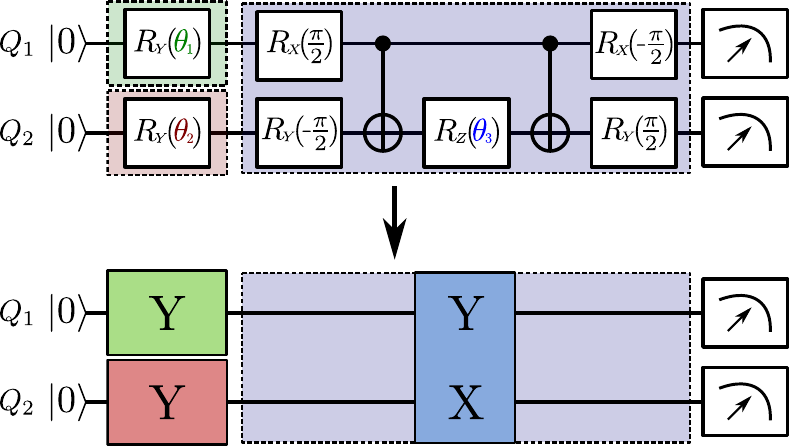}
	\caption{\label{fig:UYYX_circ}(top) A circuit to implement the YYX two-qubit variational ansatz in terms of Pauli rotations $R_A(\theta)=e^{iA_j\theta}$ on single qubits $j$ ($A=X,Y,Z$) and CNOT gates. Shaded regions denote subcircuits to implement the three separate unitary rotations in Eq.~\eqref{eq:UYYX_def}, as color coded with the variational parameters $\theta_i$. (below) The above circuit in a compressed notation, treating each rotation $U$ as a single gate labeled by the elements of the rotation generators (Eq.~\eqref{eq:generator_def}) on each qubit.}
	\end{figure}

	A quantum circuit that implements this toy ansatz is given in Fig.~\ref{fig:UYYX_circ}, using standard methods~\cite{Whitfield11} to decompose the two-qubit $e^{i\theta_2Y_1X_2}$ term in terms of single-qubit rotations and CNOT gates.

	VQEs are appealing because they reduce the computational complexity of searching the (exponentially large) $\numqubits$-qubit Hilbert space to the complexity of searching the parameter space (which may be made arbitrarily small).
	However, this comes at a cost, as none of the states $|\psi(\vec{\theta})\>$ may be close (in energy or overlap) to the target ground state.
	The variance in the energy $\<\psi|H|\psi\>$ of states $|\psi\>$ randomly drawn (i.e. with Haar measure) from an $\numqubits$-qubit Hilbert space is given by 
	\begin{equation}
		\frac{\|H\|_F^2-\mathrm{Trace}[H]^2}{4^{\numqubits}}\leq\frac{\|H-\mathrm{Trace}[H]\|_S^2}{2^{\numqubits}},
	\end{equation}
	with $\|\cdot\|_F$ the Frobenius norm and $\|\cdot\|_S$ the spectral norm.
	This implies that the probability of a random state having energy close to the ground state energy of $H$ scales as $e^{-2^{\numqubits}}$, while one expects the volume of space explored by a VQE to grow only as $e^{\numparams}$.
	This, and similar results for derivatives of the energy with respect to variational parameters~\cite{McClean18}, imply that random ansatz choice has little to no chance of success for finding ground state energies.
	Instead, a variational ansatz should be designed to cover as much of the $\numqubits$-qubit Hilbert space as possible, in a way that maximises the chance of finding low-energy states (or states that overlap well with the true ground state).

	A full VQE protocol must also concern itself with optimizing the minimization procedure, especially to prevent being stuck in local minima or barren plateaus~\cite{McClean18}.
	One should further take care to make the resulting quantum circuit as hardware efficient~\cite{Kandala17,Sagastizabal19} as possible.
	Hardware-efficiency is an active field of research and dependent upon the physical implementation of the quantum computer, and recent work has gone into optimizing the minimization procedure of a VQE~\cite{Guerreschi17,Romero18}, including the choice of cost function to minimize (e.g. to target excited states~\cite{higgott2018variational,endo2018variational}).
	In this work, we focus instead on studying the variational ansatzes themselves.
	We first focus on constructing `geometrically efficient' variational ansatzes. Then we tailor these to target specific Hamiltonians based on a perturbative approach. This generic approach is in complement with previous work on ansatz design targeting specific (classically hard) problems of interest in e.g. optimization~\cite{Farhi14} and quantum chemistry~\cite{Mcclean16}.	
	
	To pin down a working definition of `fundamentally digital' quantum ansatzes, we will use the following conditions (similar to those stated in~\cite{McClean18,Guerreschi17,Romero18,Nakanishi19}):
	\begin{dfn}
	    \label{dfn:digital_ansatz}
		A variational ansatz $(U,|\vec{0}\>)$ is a \textbf{product ansatz} if it is a product of units $U_i$,
		\begin{equation}
			U(\vec{\theta})=\prod_{i=1}^{\numunitaries}U_i(\theta_{n_i}),\label{eq:def_product_ansatz}
		\end{equation}
		where each $U_i$ has a generator $T_i$:
		\begin{equation}
		U_i(\theta_{n_i})=e^{iT_i\theta_{n_i}}.\label{eq:generator_def}
		\end{equation}
		If $n_i>n_j$ whenever $i>j$, we call the ansatz \textbf{ordered}, and if each generator is a Pauli operator - $T_i\in\PP^{\numqubits}:=\{I,X,Y,Z\}^{\otimes\numqubits}$ - we call the ansatz a \textbf{Pauli-type ansatz}.
	\end{dfn}
	We take the product in Eq.~\eqref{eq:def_product_ansatz} from right to left (i.e. $U_1(\theta_{n_1})$ acts first on the state $|\vec{0}\>$).
	As we allow $n_i=n_j$ when $i\neq j$, we may have strictly more unitaries than parameters: $\numunitaries\geq\numparams$. 
	In the rest of the text, we will refer to Pauli-type ansatzes as fundamentally digital: note that Pauli rotations can be directly implemented in a quantum circuit via the techniques of \cite{Whitfield11}. 
	When used in a VQE, Pauli-type ansatzes also have the advantage that some derivatives of the variational energy may be obtained `for free'~\cite{Nakanishi19}.
	\begin{exm}
		The YYX toy ansatz is a Pauli-type ansatz, with generators $T_1=Y_1$, $T_2=Y_2$, $T_3=Y_1X_2$.
	\end{exm}

	\subsection{Variational manifolds}\label{sec:Pauli_topology}

	Although tailoring a VQE to a Hamiltonian is essential for its success~\cite{McClean18}, interesting statements may be made about the variational ansatz prior to fixing such a target, by focusing on the manifold of states it explores.
	\begin{dfn}
		The \textbf{variational manifold} $\Mm(U,|\vec{0}\>)$ of a variational ansatz $(U,|\vec{0}\>)$ is the set $\{|\psi(\vec{\theta})\>=U(\vec{\theta})|\vec{0}\>,(\vec{\theta})\in\RR^{\numparams}\}\subset\mathbb{C}^{2^{\numqubits}}$. 
	\end{dfn}
	We note that, despite being a `manifold generated by unitary rotations', $\Mm(U,|\vec{0}\>)$ does not have a structure of a Lie group.
	This is because we only apply $U$ once to create the variational state; a state $U(\vec{\theta})U(\vec{\theta}')|\vec{0}\>$ may not correspond to any state $U(\vec{\theta}'')|\vec{0}\>$ (and most often will not).
	If $U$ is a product ansatz, one can defined a Lie group $\Ll(U)\subset \mathrm{U}(2^{\numqubits})$ from the set of generators $T_i$.
	The manifold $\Ll(U)|\vec{0}\>$ then contains $\Mm(U,|\vec{0}\>)$ as a submanifold, though it is almost always larger.
	Indeed, when $e^{i\theta T_i}$ defines a universal gate set, $\Ll(U)=\mathrm{U}(2^{\numqubits})$ and $\Ll(U)|\vec{0}\>$ is the entire set of $\numqubits$-qubit states, which is not terribly informative about the structure of $\Mm(U,|\vec{0}\>)$.

	As a rough guide, the bigger the variational manifold the better; simply adding more manifold to an ansatz can never shift it further from the target ground state.
	However, measuring the size of a variational manifold is made somewhat difficult by dimensionality concerns.
	The (real) dimension $D_{\Mm(U,|\vec{0}\>)}$ of $\Mm(U,|\vec{0}\>)$ is at most $\numparams$, but it may not achieve this upper bound, and $\Mm(U,|\vec{0}\>)$ may contain boundary regions of lower dimension.
	(Curiously, the minimal subspace of $\mathbb{C}^{2^{\numqubits}}$ containing $\Mm(U,|\vec{0}\>)$ may be of much higher dimension than $\numparams$.)
	As $\Mm(U,|\vec{0}\>)$ inherits a metric from $\mathbb{C}^{2^{\numqubits}}$, one can use this to define a Borel measure $d|\psi\>$, and thus define the area of the manifold:
	\begin{equation}
		A_{\Mm(U,|\vec{0}\>)}=\int_{\Mm(U,|\vec{0}\>)}d|\psi\>.
	\end{equation}
	When the map $(\vec{\theta})\rightarrow |\psi(\vec{\theta})\>$ is invertible on some range of parameters, its Jacobian $J$ is full-rank, and the manifold area may be calculated as
	\begin{equation}
		A_{\Mm(U,|\vec{0}\>)}=\int d^{\numparams}\theta\sqrt{\det(J^{\dag}J)}.
	\end{equation}
	However, when evaluating this integral one must take care to avoid double-counting points $\vec{\theta}\neq\vec{\theta}'$ when $|\psi(\vec{\theta})\>=|\psi(\vec{\theta}')\>$.

	\begin{exm}
		For the YYX toy ansatz, one may calculate
		\begin{equation}
			J^{\dag}J=\left(\begin{array}{ccc}1 & 0 & -\sin(2\theta_2)\\ 0 & 1 & 0 \\ -\sin(2\theta_2) & 0 & 1\end{array}\right).
		\end{equation}
		The variational manifold $\Mm(U_{YYX},|00\>)$ double-covers the Hilbert space, as
		\begin{equation}
		|\psi(\theta_3-\pi/2,\pi/2-\theta_2,\theta_1-\pi/2)\>=|\psi(\theta_3,\theta_2,\theta_1)\>
		\end{equation}
		(no other identifications exist).
		Following this identification, one can evaluate $A_{\Mm(U_{YYX},|\vec{0}\>)}=\pi^2$.
	\end{exm}
	
	\section{Stabilizer ansatzes}\label{sec:stablizer_ansatze}

	Clearly the largest space that can be spanned by any variational ansatz is the entire Hilbert space.
	The minimal number of (real) parameters required to achieve this spanning is $2(2^{\numqubits}-1)$, and it is an interesting question whether this may be provably achieved.
	In this section we answer this question in the affirmative, constructing a class of ansatzes from sequential layers of $n=1,\ldots,\numqubits$-qubit stabilizer groups~\cite{Gottesman97} (defined in App.~\ref{app:prelim}).
	Although such a construction has impractically large overhead, one may use this construction as a base to generate tractable variational ansatzes with the methods developed in Sec.~\ref{sec:child_ansatze} and Sec.~\ref{sec:hierarchies}.
	\begin{dfn}
		A \textbf{stabilizer ansatz} $(U,|\vec{0}\>)$ on $\numqubits$ qubits is constructed by choosing for each $n=1,\ldots,\numqubits$:
		\begin{enumerate}
			\item A $[n-1,n-1]$ stabilizer group $\Ss^{(n)}$, and
			\item A single-qubit starting state $|s_n\>$ for the $n$-th qubit, and
			\item Two single-qubit Pauli operators $R^{(n)}_0,R^{(n)}_1$, such that $\<s_n|R_i|s_n\>=0$, and $\mathrm{Trace}[R_0R_1]=0$.
		\end{enumerate}
		Then, one takes $|\vec{0}\>=\otimes_{n=1}^{\numqubits}|s_n\>$, and $U=\prod_{n=1}^{\numqubits}U^{(n)}$, where
		\begin{equation}
			U^{(n)}=\prod_{j=0,1}\prod_{S\in\Ss^{(n)}}e^{i\theta^n_{S,j}R^{(n)}_jS}.\label{eq:un_def}
		\end{equation}
	\end{dfn}

\begin{figure*}
		\includegraphics[width=\textwidth]{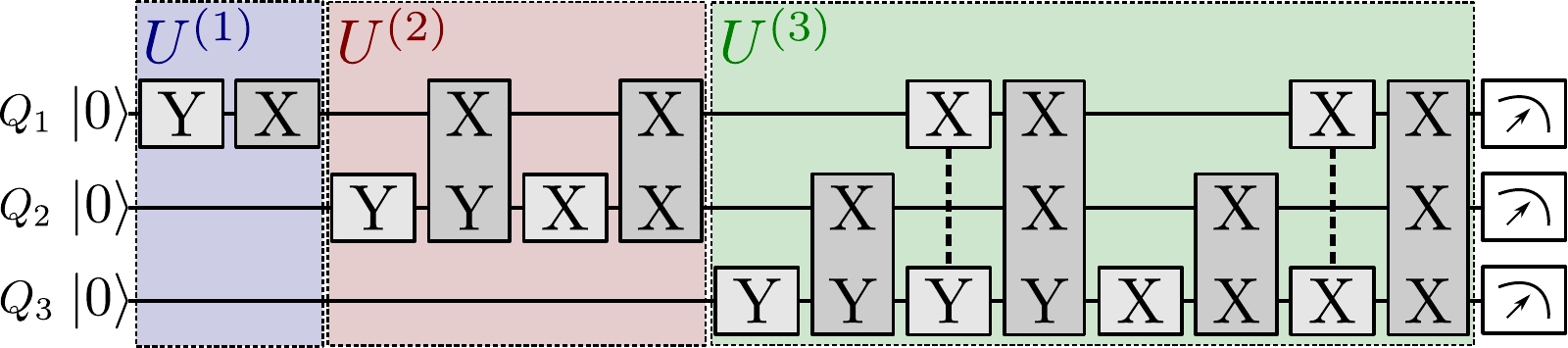}
		\caption{\label{fig:QCA3}A circuit for the QCA on $3$ qubits. For simplicity, we label each circuit element $U_i(\vec{\theta})$ by the tensor factors of its generating Pauli operator $T_i$ ($=:R^{(n)}S$ in Eq.~\eqref{eq:un_def}) on each qubit. For example, the label XXX corresponds to the rotation $e^{i\theta^3_{XX,0}XXX}$. This compression may be expanded on as shown in Fig.~\ref{fig:UYYX_circ} using the methods of~\cite{Whitfield11}. For $\numqubits$ qubits, QCA contains $2(2^\numqubits-1)$ gates and is proven to cover the entire Hilbert space (Theorem \ref{thm:minimal_spanning}). In a practical application, QCA is to be reduced to polynomial size via a hierarchical approach outlined in Sec.~\ref{sec:hierarchies}. 
		Note that the order of gate multiplication in QCA does not imply the order of gate importance in the hierarchical reduction scheme of Sec.~\ref{sec:hierarchies}. For instance, consider an application of the displayed QCA circuit to the open transverse-field Ising chain (Sec.~\ref{chapt:TFIM}). In this case, the two gates preferred in the reduction are those generated by Paulis $XYI$ and $IXY$, followed by the one generated by $XIY$ (cf. Fig.~\ref{fig:fourth_order_connected}).}
	\end{figure*}

	The definition above allows for any choice of the $[n-1,n-1]$ stabilizer groups $\Ss^{(n)}$, including ones with non-commuting elements between different $\Ss^{(n)}$.
	However, we use the following prototypical example throughout the rest of this text.
	\begin{exm}
		\label{exm:QCA}
		The \textbf{quantum combinatorial ansatz}, or \textbf{QCA}, is a stabilizer ansatz with $|s_i\>=|0\>$, $R_0^{(n)}=X$, $R_1^{(n)}=Y$, and $\Ss^{(n)}=\<X_i,i=1,\ldots,n-1\>$.
	\end{exm}
	A compressed circuit for the quantum combinatorial ansatz on $3$ qubits is given in Fig.~\ref{fig:QCA3}

	\begin{thm}
	\label{thm:minimal_spanning}
		A stabilizer ansatz $(U,|\vec{0}\>)$ spans the entire Hilbert space of $\numqubits$-qubit states with the minimal number of parameters.
	\end{thm}
	\emph{Proof ---} That the number of parameters is minimal may be immediately calculated, 
	\begin{equation}
	\numparams=\sum_{n=1}^{\numqubits}2\times 2^{n-1}=2(2^{\numqubits}-1).
	\end{equation}
	We then prove that the ansatz spans the entire Hilbert space by induction.
	The stabilizer group $\Ss^{(n)}$ gives a basis $|p\>$ for the $n-1$ qubit Hilbert space.
	Then, as $[R^{(n)}_jS,R^{(n)}_jS']=0$, one may rewrite $U^{(n)}$ as
	\begin{equation}
		U^{(n)}=\prod_{j=0,1}\exp\left[i\sum_{S\in\Ss^{(n)}}\theta^n_{S,j}R^{(n)}_jS\right].
	\end{equation}
	This sends the state $|p\>|s_n\>$ to the state
	\begin{equation}
		|p\>\left(e^{i\theta^{n}_{p,0}R^{(n)}_0}e^{i\theta^n_{p,1}R^{(n)}_1}\right)|s_n\>,
	\end{equation}
	where the angles $\theta^n_{p,j}$ are given by the following linear transformation:
	\begin{equation}
		\theta^n_{p,j}=\sum_{S\in\Ss^{(n)}}S_p\theta^n_{S,j},\hspace{0.5cm} S_p=\<p|S|p\>\in\{\pm 1\}.
	\end{equation}
	This is the Hadamard-Walsh transformation, which is invertible, so $\theta^n_{p,j}$ can now be treated as independent parameters. On the other hand, our choice of $R^{(n)}_j$ explicitly takes the starting state $|s_n\>$ on qubit $n$ to any state on the Bloch sphere.
	This implies that if we have the ability to create an arbitrary $n-1$-qubit state
	\begin{equation}
		|\Psi^{(n-1)}\>=\sum_pa_p|p\>,
	\end{equation}
	$U^{(n)}|\Psi^{(n-1)}\>|s_n\>$ may be tuned to achieve any state of the form
	\begin{equation}
	\sum_{p}a_p|p\>\left(e^{i\theta^{n}_{p,0}R^{(n)}_0}e^{i\theta^n_{p,1}R^{(n)}_1}\right)|s_n\>,
	\end{equation}
	which describes an arbitrary $n$-qubit state.
	This then completes the proof of coverage by induction, as $U^{(1)}|s_1\>$ covers the entire Bloch sphere.\qed
	
	\section{Children ansatzes and their construction}\label{sec:child_ansatze}
	The cost of implementing a product VQE grows polynomially in both the number of units $\numunitaries$ (as this dictates the circuit size) and the number of parameters $\numparams$ (as this dictates the size of the optimization problem).
	Thus, an ansatz that covers the entire Hilbert space is too expensive to be of use; one must use it to construct child ansatzes of a manageable size.
	\begin{dfn}
		A product ansatz $(U',|\vec{0}'\>)$ is a \textbf{child ansatz} of a \textbf{parent} product ansatz $(U,|\vec{0}\>)$ when each unit $U'_i$ of $U'$ also appears in $U$.
	\end{dfn}
	This definition is operational rather than fundamental; the variational manifold of a child ansatz is not necessarily a submanifold of the parent ansatz' variational manifold.
	However, one expects that these children ansatzes will still inherit some properties of the parent.
	In particular, we expect that a parent ansatz that spans as large a part of the Hilbert space as possible will lead to children ansatzes that are similarly large.
	
	\subsection{Ansatz compression and hierarchical construction \label{sec:compression_hierarchy}}

	An obvious method to construct a child ansatz from a parent is to simply get rid of individual units or parameters:
	\begin{dfn}
		Given a product ansatz $(\prod_jU_j(\theta_{n_j}),|\vec{0}\>)$, one may \textbf{remove} a parameter $\theta_{n_i}$ to obtain the child ansatz $(\prod_{n_j\neq n_i}U_j(\theta_{n_j}),|\vec{0}\>)$, or \textbf{fix} a parameter $\theta_{n_i}=c\theta_{n_j}$ with $c\in\RR$ to obtain the child ansatz $(\prod_{l}U'_l(\theta_{m_l})|\vec{0}\>)$, where $m_i=n_j$, $m_l=n_l$ for $l\neq i$, and $T'_l=cT_l$ whenever $n_l=n_i$.
	\end{dfn}
	Parameter fixing may be considered strictly more general than unit removal, as fixing $\theta_{n_i}=0\theta_{n_j}$ produces the same variational manifold as removing $\theta_{n_i}$.
	However, unit removal reduces both $\numunitaries$ and $\numparams$, while parameter fixing does not reduce the resulting circuit length.

	Alternatively, one may construct child ansatzes using a bottom-up approach:
	\begin{dfn}
		Given a product ansatz $(\prod_jU_j(\theta_{n_j}),|\vec{0}\>)$, one may construct a \textbf{priority list} $(U_{j_1},U_{j_2},\ldots)$ of the possibly-repeated units of the ansatz.
		Such a priority list allows the construction of a \textbf{hierarchy} of child ansatzes $(U_M,|\vec{0}\>)$ (for $M>0$), where
		\begin{equation}
			U_M(\vec{\theta})=\prod_{m=1}^MU_{j_m}(\theta_{n_m}).
		\end{equation}
	\end{dfn}
	The two methods described above may be combined if desired.
	Subsequent generations of ansatzes will trade off a lower cost to implement against a smaller-sized variational manifold.
	We now focus on methods to optimize this balance.
	We first demonstrate how one may use unit reduction and parameter fixing to force a large VQE to respect symmetry constraints on the system.
	Following this, we take a rigorous perturbative approach to construct priority lists for a given target Hamiltonian.
	
	\subsection{Compression over symmetries \label{sec:symmetry_enforcing}}
	One may often restrict the ground state of a system by symmetries of the Hamiltonian; that is, operators $S$ that commute with $H$.
	When this is true, all eigenstates $|E_0\rangle$ of $H$ may be chosen to be eigenstates of $S$.
	This is particularly relevant in electronic systems where the particle number $\sum_iZ_i$ or parity $\prod_iZ_i$ is conserved.
	The symmetry is enforced on all states in a variational ansatz $(U, |\vec{0}\>)$ when $|\vec{0}\>$ is an eigenstate of $S$, and $[U(\vec{\theta}),S]=0$ for all choices of the parameters $\theta$.
	This in turn requires for an ordered product ansatz $U(\vec{\theta})=\prod_iU_i(\theta_{n_i})$ that $[\prod_{i,n_i=n}U_i(\theta_{n_i}),S]=0$ for all unique parameters $n$ and for all choices of $\theta_{n_i}$.
	If a parameter $\theta_{n_i}$ is associated to a single generator $T_i$, then this occurs if and only if $[T_i,S]=0$.

	When a symmetry is not respected by a variational ansatz, one may choose to either remove or fix the offending terms (see \cite{Gard20} for an alternative approach).
	Removal of generators that do not respect a given symmetry is simplest, but may be too restrictive for our desires.
	One may fix an ordered product ansatz to obey a symmetry that is broken by a set of commuting generators $\{T_{M_0},T_{M_0+1},\ldots,T_{M_1}\}$.
	To do this, one needs to solve the system of linear equations
	\begin{equation}
		\sum_{m=M_0}^{M_1} c_m\sum_{i,n_i=m}[S,T_i] = 0,
	\end{equation}
	and fix $c_n\theta_n=c_m\theta_n$ for $N\leq n,m\leq M$.
	This requires fixing \emph{all} parameters between $N$ and $M$, which in turn might require rearranging the original ansatz to place specific units next to each other.

	A very simple symmetry to enforce in a problem is the (antiunitary) complex conjugation operator, $\Kk i=-i\Kk$.
	(This symmetry is respected whenever the Hamiltonian is purely real.)
	As we have defined our generators $T_i$ with an imaginary unit, $U_i=e^{i\theta_{n_i}T_i}$ commutes with $\Kk$ when $T_i$ anti-commutes with $\Kk$.
	(e.g.~for a single qubit, the rotation $e^{i\theta Y}$ rotates between the real eigenstates of the real $X$ and $Z$ Pauli operators.)
	\begin{exm}
		The YYX toy ansatz is the compression of the QCA stabilizer ansatz for two qubits over $\Kk$. It thus spans the entire Hilbert space of 2-qubit states with real coefficients (which matches the calculation of its variational area).
	\end{exm}

	In App.~\ref{app:UCC}, we give another example of a symmetry-compressed Pauli-type ansatz - the fermionic unitary coupled cluster ansatz.
	
	\subsection{Size-extensivity of a variational ansatz}\label{sec:size_extensive}
	
	To show beyond-classical performance, we desire our variational quantum algorithms to be able to produce strongly entangled states, inaccessible to a classical computer.
	For this, we would like the VQE ansatz to represent quantum correlations in a maximally compact manner.
	To achieve this, we are guided by the idea of size-extensivity.
	The notion of size-extensivity has its origins in strongly-correlated physics, and is formalized there by the linked cluster theorem~\cite{brueckner55}.
	The rough notion is that: (1) if a computation treats two uncoupled systems together, it should converge to the same solution as when it treats them independently, and (2) the only complexity one should be adding to the solution of coupled systems is that which is minimally demanded.
	Formalizing this idea requires somewhat heavy machinery; we give a formal definition later in the text (Def.~\ref{dfn:size_extensive}) and now put forward the following (weaker) statement as an informal definition.

	\begin{dfn}\label{dfn:size_extensive_informal}(informal) Consider variational ansatz $(U,|\vec{0}\rangle)$ for a Hamiltonian $H$ on a system $S$, and an arbitrary (disjoint) partition $S=\cup_i S_i$ with a decomposition $H=\sum_i H_i+H_{\mathrm{other}}$ where each $H_i$ acts only on $S_i$ (and each term in $H_{\mathrm{other}}$ acts on multiple $S_i$). In this case, the ansatz $(U,|\vec{0}\rangle)$ is size-extensive if for any such partition, the unitary $U(\vec{\theta})$ that minimizes the variational energy $E(\vec{\theta})$ (Def.~\ref{dfn:var_ansatz}) reduces to the form
	    \begin{equation}
	        U(\vec{\theta})=\prod_i U(\vec{\theta}_i)
	        \label{eq:size_extensive_informal}
	    \end{equation}
	if $H_{\mathrm{other}}$ is reduced to 0. In \eqref{eq:size_extensive_informal}, each $U(\vec{\theta}_i)$ acts only on system $S_i$ (i.e. the coefficients of any part of the ansatz $U$ that acts outside of $S_i$ are set to $0$).
	\end{dfn}
	
	In the language of Def.~\ref{dfn:size_extensive_informal}, the stronger statement of Def.~\ref{dfn:size_extensive} is needed to treat the case where $\{S_i\}$ together form a connected system, but some pairs $(S_k, S_l)$ are mutually separated (e.g. because of spatial locality). It appears that in this case, a variational ansatz is efficient if it tends to introduce more correlations between less separated subsystem pairs $(S_k, S_l)$. However, this heuristic needs to be re-stated more rigorously. In Def.~\ref{dfn:size_extensive}, we provide such a rigorous formulation and apply it in an explicit construction of size-extensive ansatzes.

	\section{Perturbative construction for digital size-extensive ansatzes}\label{sec:hierarchies}
	
	We now propose a perturbative approach for the construction of digital size-extensive ansatzes. We formulate it in terms of a gate hierarchy list $(U_1,\ldots)$ derived from a large parent ansatz $(U,|\vec{0}\>)$. To decide on the hierarchy list, we split the system Hamiltonian $H$ into the non-interacting part $H_0$ and the coupling $J V$ ($\|H_0\|,\|V\|\sim 1$): 
	\begin{equation}
	H=H_0 + J V\label{eq:PT_ham}
	\end{equation}
	To allow for analytical treatment we consider the weak coupling limit, $J\ll1$.
	In this limit, the overlap between the true ground state $|E_0\>$ and unperturbed excited states $|E_j^0\>$ is exponentially small in the number of applications of $V$ required to couple $|E_j^0\>$ to the unperturbed ground state $|E_0^0\>$.
	We may rewrite the non-interacting part $H_0$ via a unitary transformation as
	\begin{equation}
	H_0=\sum_{n=1}^{\numqubits}h_nZ_n,
	\label{eq:unperturbed_hamiltonian}
	\end{equation}
	which ties each $|E_j^0\>$ to a computational basis state $|\vec{s}\>$
	\begin{equation}
	H_0|\vec{s}\rangle=-\sum_{n=1}^{\numqubits}(-1)^{s_n}h_n|\vec{s}\rangle.
	\end{equation}
	If we can further tie each state $|\vec{s}\>$ to one or a few variational units $U_i(\theta_i)$, we can construct a hierarchy list of these $U_i(\theta_i)$ based on the approximate magnitude of $|\<\vec{s}|E_0\>|$. The resulting hierarchy list is to be used in the VQE procedure for the original, potentially strongly coupled Hamiltonian $H$ ($J=O(1)$).
	
	Performing this construction in a size-extensive way runs into a challenge which we call `back-action'.
	Namely, the action of any unit $U_i(\theta_i)$ on the state $\prod_{j<i}U_j(\theta_j)|\vec{0}\>$ may be very different to the action of $U_i(\theta_i)$ on the starting state.
	In particular, one could imagine this action generating an undesired term to the variational wavefunction which must be cancelled by later rotations.
	As we will show, one can deal with this back-action while retaining the size-extensivity. To achieve this, in the rest of this section we will expand the target equality,
	\begin{equation}
	\ket{E_0}\simeq \ket{\psi(\vec{\theta})}, \label{QCA_approximation}\\
	\end{equation}
	assuming that $\ket{\psi(\vec{\theta})}$ is given by a digital (i.e., Pauli-type) ansatz. We will do so in terms of a Pauli decomposition of the perturbation 
	\begin{equation}
	JV=\sum_{i=1}^{N_c}J_i V_i,\hspace{1cm}V_i\in\PP^{\numqubits},
	\label{eq:perturbation}
	\end{equation}
	and then we will equate terms based on the order of their polynomial dependence on each $J_i$.
	On the left-hand side (Sec.~\ref{sec:digital_PT}), we will use a Dyson expansion, with a convenient diagrammatic representation. On the right-hand side (Sec.~\ref{sec:taylor}) we will use a Taylor expansion of the exponential operators.
	We will show that a single condition (Def.~\ref{dfn:generating}) on the parent ansatz is sufficient to automatically cancel all undesired back-action.
	Then, we will show that an additional condition (Def.~\ref{dfn:matched}) causes the back-action terms to precisely cancel out any need for entangling circuits between disconnected regions (Theorem \ref{thm:no_disconnected_diagrams}). This ensures the desired feature of size-extensivity, thus providing the digital quantum version of the linked-cluster theorem \cite{brueckner55}.
	The QCA ansatz of Example~\ref{exm:QCA} will be seen to satisfy the above conditions, and therefore gives rise to a hierarchy of size-extensive digital ansatzes.

	\par
	Our perturbative approach can be thought of as a digital unitary relative of the Kirkwood-Thomas expansion~\cite{KT,BKL}. Also note, that as we intend to optimize the parameters $\vec{\theta}$ as part of the VQE, we will approximate these only to leading order in the interaction strength J. This makes our method potentially applicable even in the strongly correlated regime where perturbation theory breaks down.
	\subsection{Diagrammatic expansion of the ground state \label{sec:digital_PT}}
	To expand the left-hand side of Eq.~\eqref{QCA_approximation}, let us use vector notation $\vec{J}$ for the coupling terms $J_i$ (and $\vec{V}$ for the operators $V_i$).
	Then, let us introduce some notation that simplifies the following expressions:
	\begin{equation}
	\vec{a}^{\cdot\vec{k}}:=\prod_ia_i^{k_i}=\exp(\vec{k}\cdot\log(\vec{a})).\label{eq:vectorpowernotation}
	\end{equation}
	We wish to use this expression for both vectors of numbers (e.g.~$\vec{J}$) and vectors of operators (e.g.~$\vec{V}$).
	In the latter we must take care of ordering; as previous, we assume that the product runs right-to-left.
	As Pauli operators either commute or anticommute, rearranging these products simply requires one to keep track of minus signs.
	This may be assisted by the following definition
	\begin{dfn}
	A vector $\vec{V}$ of $\numcouplings$ Pauli operators defines a phase $\Gamma(\vec{k})\in\{0,1,2,3\}$ and a state $\vec{s}(\vec{k})$ on a vector $\vec{k}\in\mathbb{N}^{\numcouplings}$~\footnote{We take the natural numbers $\mathbb{N}$ to include $0$.} by
	\begin{equation}
	\vec{V}^{\cdot\vec{k}}|\vec{0}\rangle=i^{\Gamma(\vec{k})}|\vec{s}(\vec{k})\rangle,
	\label{eq:def_s(k)_phase(k)}
	\end{equation}
	and a relative sign $S_{\vec{k},\vec{k}'}\in\{-1,1\}$ for $\vec{k},\vec{k}'\in\mathbb{N}^{\numcouplings}$ by
	\begin{equation}
	\vec{V}^{\cdot\vec{k}}\vec{V}^{\cdot\vec{k}'}=S_{\vec{k},\vec{k}'}\vec{V}^{\cdot(\vec{k}+\vec{k}')}.\label{eq:def_relative_sign}
	\end{equation}
	\end{dfn}
	Then, as Pauli operators map computational basis states to computational basis states, $\vec{V}^{\vec{k}}|\vec{0}\>$ is an eigenstate of $H_0$, with energy
	\begin{equation}
	E_{\vec{s}(\vec{k})}:=-\sum_{n=1}^{\numqubits}(-1)^{\vec{s}(k)_n}h_n.
	\end{equation}

	Let us now expand the ground state as a Taylor series in $\vec{J}$:
	\begin{equation}
	|E_0\rangle = \sum_{\vec{k}\in\mathbb{N}^{\numcouplings}}\vec{J}^{\cdot\vec{k}}|\Psi_{\vec{k}}\rangle.\label{eq:PT_expansion}
	\end{equation}
	Following a standard Dyson expansion (details in App.~\ref{app:proof_k_determines_s}), we observe that
	\begin{lem}
	\label{lem:Psi_k_expression}
	The vectors $|\Psi_{\vec{k}}\rangle$ take the form
	\begin{equation}
	|\Psi_{\vec{k}}\rangle=C_{\vec{k}}\vec{V}^{\cdot\vec{k}}|\vec{0}\rangle,\label{eq:Psi_k_expression}
	\end{equation}
	where $C_{\vec{k}}$ is a real number.
	\end{lem}
	To find the values of coefficients $C_{\vec{k}}$, we first develop a perturbative expansion for a ground state $\ket{\tilde{E}_0}$ with a special normalization condition $\braket{\vec{0}|\tilde{E}_0}=1$,
	\begin{equation}
	|\tilde{E}_0\rangle = \sum_{\vec{k}\in\mathbb{N}^{\numcouplings}}\vec{J}^{\cdot\vec{k}}|\tilde{\Psi}_{\vec{k}}\rangle.\label{eq:PT_tilde_expansion}
	\end{equation}
	The states $|\tilde{\Psi}_{\vec{k}}\rangle$ then satisfy (see App.~\ref{app:proof_k_determines_s}):
	\begin{equation}
	|\tilde{\Psi}_{\vec{k}}\rangle=\tilde{C}_{\vec{k}}\vec{V}^{\cdot\vec{k}}|\vec{0}\rangle,\label{eq:Psi_tilde_k_expression}
	\end{equation}
	where $\tilde{C}_{\vec{k}}$ is a real number.
	In particular, if $\vec{\delta}_{\beta}$ is the unit vector with a $1$ in the $\beta$ index, $\tilde{C}_{\vec{k}}=\delta_{\vec{k},\vec{0}}$ if $\vec{s}(\vec{k})=\vec{0}$, and is otherwise given by the recursive relation
	\begin{align}
	\tilde{C}_{\vec{k}}=&(E_{\vec{0}}^{(0)}-E_{\vec{s}(\vec{k})}^{(0)})^{-1}\sum_{\beta,k_{\beta}>0}\left\{\tilde{C}_{\vec{k}-\vec{\delta}_{\beta}}S_{\vec{\delta}_{\beta},\vec{k}-\vec{\delta}_{\beta}}\nonumber\right.\\
	&-\sum_{\substack{\vec{k}'<\vec{k},\;k'_{\beta}>0\\\vec{s}(\vec{k'})=0}}\left.\tilde{C}_{\vec{k}'-\vec{\delta}_{\beta}}\tilde{C}_{\vec{k}-\vec{k}'}S_{\vec{\delta}_{\beta},\vec{k}'-\vec{\delta}_{\beta}}S_{\vec{k}-\vec{k}',\vec{k}'}\right\},\label{eq:tilde_Ck}
	\end{align}
	where $\vec{k}'<\vec{k}$ if $k'_{\beta}\leq k_{\beta}$ for all $\beta$ and $\vec{k}'\neq\vec{k}$.
	To find the coefficients $C_{\vec{k}}$ of the normalized ground state, one may then expand the expression $\ket{E_0}=\braket{\tilde{E}_0|\tilde{E}_0}^{-1/2}\ket{\tilde{E}_0}$ in powers of $\vec{J}$, which allows to express $C_{\vec{k}}$ in terms of $\tilde{C}_{\vec{k}}$ obtained from \eqref{eq:tilde_Ck}.
	
	We note here that we have no guarantee that the normalization constant $\mathcal{N}=\braket{\tilde{E}_0|\tilde{E}_0}^{-1/2}$ behaves regularly in thermodynamic limit $\numqubits\rightarrow\infty$.
	This is a standard breakdown of perturbation theory for the wavefunction, however when this occurs our approach to VQE construction is still possible, and may indeed still be practical.
	At the stage of estimating the variational parameters $\vec{\theta}$, we will be using the $\tilde{C}_{\vec{k}}$ coefficients, since they behave regularly and are more practical to calculate.
	As $\vec{\theta}$ will be optimized later on the quantum device, the estimation itself need not be exact.

	The size-extensivity of our approach relies on an important relationship between $C_{\vec{k}}$ terms that are the combination of disconnected pieces.
	To formalize this notion of connectedness, we introduce some terminology:
	\begin{dfn}
		For a perturbative contribution $C_{\vec{k}}$, the set of couplings $V_i$ s.t. $k_i\neq0$, is said to be \textbf{activated} in $\vec{k}$.
		The set of qubits on which at least one activated coupling $V_i$ acts non-trivially is called the \textbf{support} of $\vec{k}$.
		\label{dfn:support}
	\end{dfn}
	Then the connectedness of the contribution $C_{\vec{k}}$ is defined as follows:
	\begin{dfn}
		\label{dfn:disconnected_contributions}
		A perturbative contribution $C_{\vec{k}}$ is disconnected if one may write
		\begin{equation}
		\vec{k}=\vec{k}_A+\vec{k}_B, \label{eq:disconnected_k_subtle}
		\end{equation}
		such that the respective supports of $\vec{k}_A$ and $\vec{k}_B$ do not share any qubits.
		This implies, but is not equivalent to, the following statement:
		\begin{equation}
		\label{eq:disconnected_k}
		\vec{V}^{\cdot\vec{k}}=\vec{V}^{\cdot\vec{k}_A}\vec{V}^{\cdot\vec{k}_B}
		\end{equation}		
	\end{dfn}
	The disconnected contributions $C_{\vec{k}}$ obey the following special property (proven in App.~\ref{app:lem_disconnected_contributions_proof}).
	\begin{lem}
		\label{lem:disconnected_contributions}
		If a perturbative contribution $C_{\vec{k}}$ is disconnected w.r.t. a splitting \eqref{eq:disconnected_k} into $\vec{k}_A$ and $\vec{k}_B$,
		\begin{equation}
		C_{\vec{k}}=C_{\vec{k}_A}C_{\vec{k}_B}.
		\end{equation}		
	\end{lem}

\begin{figure}
	\includegraphics[width=\columnwidth]{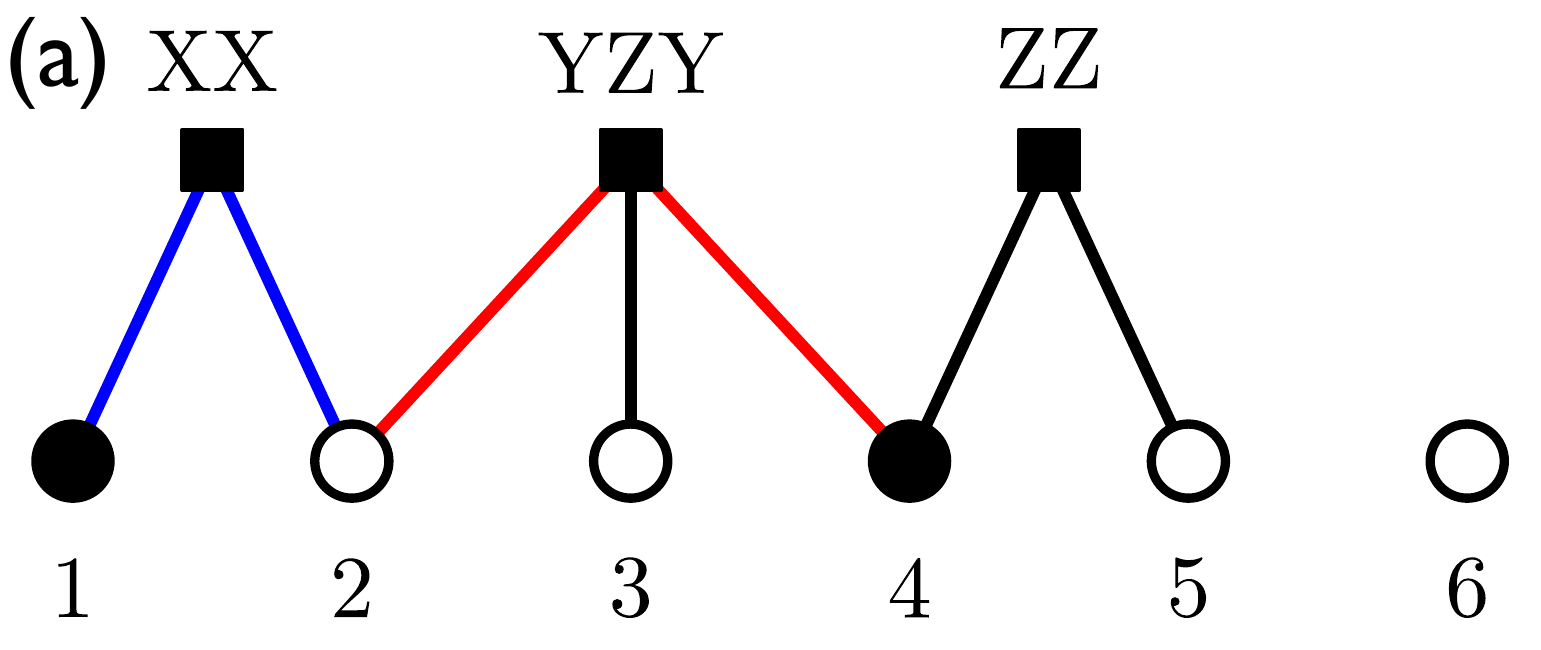}\\
	\includegraphics[width=\columnwidth]{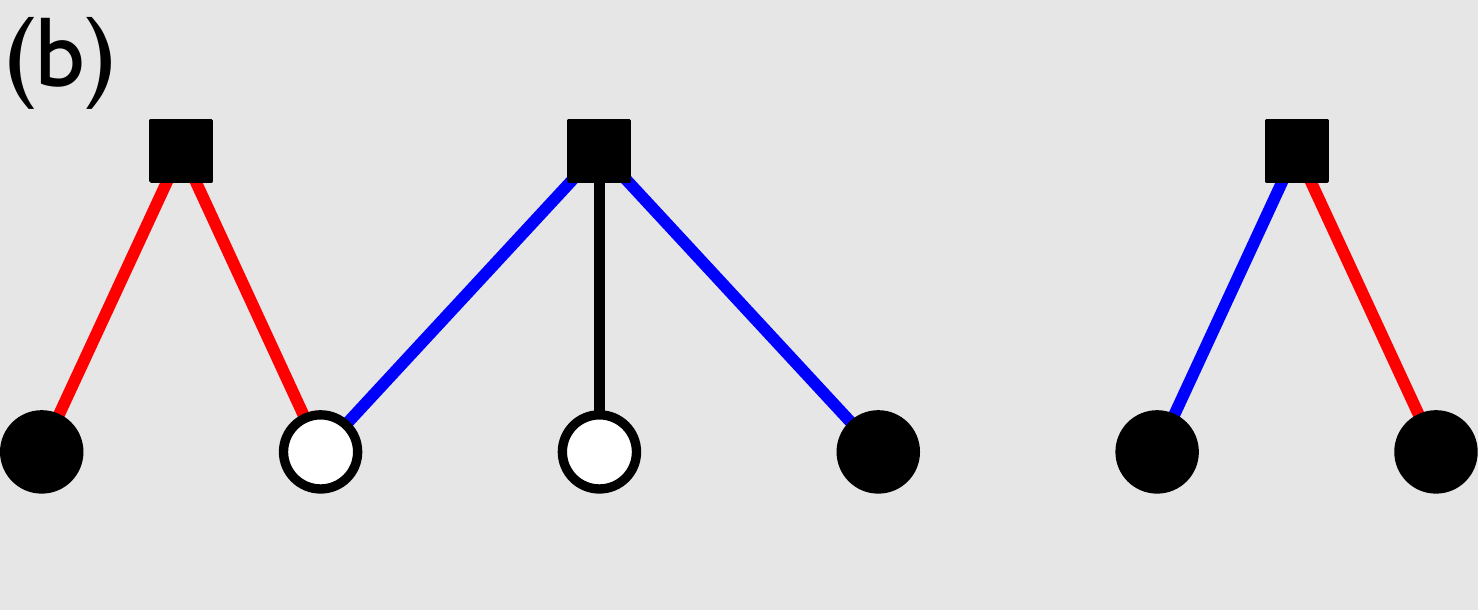}
	\caption{\label{fig:example_diagrams_new}Example perturbative diagrams. (a) A connected diagram for a real contribution (even number of $Y$ terms) to $|\vec{s}(\vec{k})\rangle=|100100\rangle$. Labels for qubits $i$ and terms $V_{\beta}$ are added for reference. (b) A disconnected diagram for an imaginary contribution to $|\vec{s}(\vec{k})\rangle=|100111\rangle$. Unnecessary labels here are excluded.}
	\end{figure}

	This idea of connectedness of contributions may be described in a graphical representation of the product of operators $\vec{V}^{\cdot\vec{k}}$:
	\begin{dfn}
		Let $\vec{V}$ define the order of a decomposition of the perturbation $\vec{J}\cdot\vec{V}$ to a non-interacting Hamiltonian $H_0$.
		A perturbative diagram for a vector $\vec{k}$, is a bipartite graph with one circular vertex for each qubit, and $k_{\beta}$ square vertices for each interaction $V_{\beta}$.
		We draw edges between each square vertex and the qubits that the corresponding $V_{\beta}$ term acts non-trivially on, and color the edge to qubit $i$ blue, red or black if $[V_{\beta}]_i=X,Y$ or $Z$ respectively.
		Each circular vertex is then coloured black or white if it is connected to by an odd or even number of coloured edges respectively.
	\end{dfn}
	A contribution $C_{\vec{k}}$ is connected if all square vertices in the perturbative diagram are connected~\footnote{The circular vertices, corresponding to qubits, need not be connected, as a connected contribution need not act on all qubits.}.
	In Fig.~\ref{fig:example_diagrams_new}, we show some examples of connected and disconnected perturbative diagrams.
	Diagrams also allow one to read off $\vec{s}(\vec{k})$ ($s_i(\vec{k})=0$ when the corresponding vertex is white), and $\Gamma(\vec{k})$ mod 2 (being the number of red lines modulo 2).
	(The rest of $\Gamma(\vec{k})$ depends on the order in which the operations $V_i$ are applied, which is not captured in the perturbative diagrams.)

	\subsection{Taylor expansion of the variational ansatz}\label{sec:taylor}
	We now consider the expansion of the right hand side of Eq.~\eqref{QCA_approximation}.
	In keeping with the previous subsection, we wish to do this in terms of the individual perturbations $J_i$.
	Let us expand each coefficient $\theta_i$ in a power series over all interaction terms $J_i$
	\begin{equation}
	\theta_i=\sum_{\vec{k}}\theta_i^{(\vec{k})}\vec{J}^{\cdot\vec{k}}\left(=\sum_{\vec{k}}\theta_i^{(\vec{k})}J_1^{k_1}J_2^{k_2}\ldots\right),
	\end{equation}
	where the shorthand vector power notation was defined in Eq.~\eqref{eq:vectorpowernotation}.
	This may be substituted into the variational ansatz $(U,|\vec{0}\>)$
	\begin{align}
	\label{eq:product_ansatz_absorbed_i}
	U(\vec{\theta})&=\prod_i\left\{\prod_{\vec{k}}\exp\left[i\theta_i^{(\vec{k})}\vec{J}^{\cdot \vec{k}}T_i\right]\right\},
	\end{align}
	where we added the brackets to emphasize the ordering of the product over $i$. 
	Now, we take the Taylor series of the exponentials in Eq.~\eqref{eq:product_ansatz_absorbed_i}, obtaining
	\begin{align}
	&U(\vec{\theta})=\prod_i\left\{\prod_{\vec{k}}\sum_{f=0}^{\infty}\frac{1}{f!}\left[i\theta_i^{(\vec{k})}\vec{J}^{\cdot \vec{k}}T_i\right]^f\right\}.\label{eq:taylorexp}
	\end{align}
	
	We will eventually wish to rearrange this product to identify all terms that share the same power of each $J_i$ --- that is, those that share the same $\vec{J}^{\cdot\vec{k}}$.
	This requires first expanding our product over sums to a sum over products (pulling the sum over integers $g$ in front of the products over $k$ and $i$).
	Each term in the resulting sum will have a unique product of powers of the different $\theta_i^{(k)}$.
	We can then associate this term to a function $\vec{f}:\mathbb{N}^{\numcouplings}\rightarrow\mathbb{N}^{\numparams}$; i.e. the power of $\theta_i^{(k)}$ in our term is given by $f_i(\vec{k})$.
	(Each such function $\vec{f}$ will correspond in Sec.~\ref{sec:equating_terms} to a unique way to map the activations of couplings $V_\alpha$ from the left-hand side of Eq.~\eqref{QCA_approximation} onto the generators $T_i$.)
	One may confirm that every unique function $\vec{f}$ corresponds to a single term in Eq.~\eqref{eq:taylorexp}, and the powers of the $T_i$, $\vec{J}$, and the coefficient of each term may be expressed in terms of this function, allowing us to expand our unitary $U(\vec{\theta})$ as
	\begin{align}
	\label{eq:expanded_U_raw}
	U(\vec{\theta})=\sum_{\vec{f}}\vec{J}^{\cdot \sum_{i,\vec{k}}f_i(\vec{k})\vec{k}}\left(i\vec{T}\right)^{\cdot\sum_{\vec{k}}\vec{f}(\vec{k})}\prod_{\vec{k},i}\frac{\left[\theta_i^{(\vec{k})}\right]^{f_i(\vec{k})}}{f_i(\vec{k})!},
	\end{align}
	To put \eqref{eq:expanded_U_raw} in a simpler form, we define:
	\begin{align}
	\vec{K}(\vec{f}) &= \sum_{i,\vec{k}'}f_i(\vec{k}')\vec{k}' \label{eq:def_K(f)}\\
	\vec{N}(\vec{f}) &= \sum_{\vec{k}}\vec{f}(\vec{k}), \label{eq:def_N(f)}\\
	\Theta(\vec{f}) &= \prod_{\vec{k},i}\frac{\left[\theta_i^{(\vec{k})}\right]^{f_i(\vec{k})}}{f_i(\vec{k})!},
	\end{align}
	which allows us to rewrite the sum as
	\begin{equation}
	\label{eq:expanded_U}
	U(\vec{\theta}) = \sum_{f:\mathbb{N}^{\numcouplings}\rightarrow\NN^{\numparams}}\vec{J}^{\cdot\vec{K}(\vec{f})}\left(i\vec{T}\right)^{\cdot\vec{N}(\vec{f})}\Theta(\vec{f}).
	\end{equation}
	One can give an interpretation for $\vec{K}(\vec{f})$, $\vec{N}(\vec{f})$ and $\Theta(\vec{f})$ in the expression. The vector $\vec{K}(\vec{f}) \in \NN^{\numcouplings}$ represents the PT order of a given term of the sum.
	(Note that multiple functions $\vec{f}$ will have the same PT order $\vec{K}(\vec{f})$.)
	$\vec{N}(\vec{f}) \in \NN^{\numparams}$ gives the activation of generators $T_i$ in that term, and therefore tells us the computational basis state that this term produces as an operator acting on $\ket{\vec{0}}$.
	(Terms with $|\vec{N}(\vec{f})|\geq 2$ describe the `back-action' of the ansatz which we will discuss in the next section.)

	The information about the parameters $\vec{\theta}$ of the ansatz is now contained in the scalar coefficient $\Theta(\vec{f})$.
	Its values are not independent variables: $\Theta(\vec{f})$ can be fixed entirely by its action on the functions $f$ s.t. $|\vec{N}(\vec{f})|=1$. To see this, let us label such functions $\vec{f}=d^{\vec{k},i}$, where $d^{\vec{k},i}_j(\vec{k}')=\delta_{\vec{k},\vec{k}'}\delta_{i,j}$. These functions yield an activation of a single generator $T_i$ from a single activation pattern $\vec{k}$ of couplings $V_\alpha$.
	For such functions, we obtain $\Theta(d^{\vec{k},i})=\theta_i^{(\vec{k})}$ -- whose values indeed entirely determine the ansatz state. In particular, in the terms describing back-action ($f$ s.t. $|\vec{N}(\vec{f})|\geq2$), $\Theta(\vec{f})$ are nonlinear monomials of $\theta_i^{(\vec{k})}$, and thus are fixed by the values of $\Theta(d^{\vec{k},i})$.

	\subsection{Equating ansatz and perturbative terms}\label{sec:equating_terms}
	
	Our plan is now to solve for $\theta_i^{(\vec{k})}$, by comparing $|\psi(\vec{\theta})\rangle$ from Eq.~\eqref{eq:expanded_U} to the perturbative series for $\ket{\Psi(\vec{J})}$ from Eq.~\eqref{eq:PT_expansion}.
	We will equate the contributions coming from different PT orders, and those proportional to the same computational basis state.
	(The vectors $\vec{K}(\vec{f})$ and $\vec{N}(\vec{f})$ allow us to identify which terms need be equated.)
	This will result in equations that are linear in the coefficients $C_{\vec{k}}$ and $\Theta(\vec{f})$.
	Due to the structure of $\Theta(\vec{f})$ these equations will be highly nonlinear in $\theta_i^{(\vec{k})}$.
	However, under certain conditions (Def.~\ref{dfn:generating} and Def.~\ref{dfn:matched}), we find that these equations for $\theta_i^{(\vec{k})}$ may be solved iteratively, and that many coefficients will vanish exactly.
	This will yield a class of ansatzes which are also size-extensive, the technical definition of which we give in Def.~\ref{dfn:size_extensive}.
	For such ansatzes, we will have a guarantee that a relatively compact circuit is capable of reproducing the perturbative series for $\ket{\Psi(\vec{J})}$ up to a given PT order $\vec{k}$.
	These circuits will have a relatively small (polynomial in $\numqubits$ at fixed PT order $\vec{k}$) number of free parameters when used as a VQE, as this coincides with the number of leading order connected diagrams up to order $\vec{k}$.

	Equating the action of the Taylor-expanded $U(\vec{\theta})$ (Eq.~\eqref{eq:expanded_U}) on the starting state $|\vec{0}\>$ to the expansion of the ground state $|E_0\rangle$ (Eq.~\eqref{eq:PT_expansion}) and separating in orders of $\vec{J}$ obtains the form
	\begin{equation}
	C_{\vec{k}}\vec{V}^{\cdot\vec{k}}|\vec{0}\rangle-\sum_{f;\vec{K}(\vec{f})=\vec{k}}\Theta(\vec{f})\left(i\vec{T}\right)^{\cdot\vec{N}(\vec{f})}|\vec{0}\rangle=0.
	\end{equation}
	This may be further separated by taking the inner product with different computational basis states to give the equations
	\begin{align}
	C_{\vec{k}}-\sum_{f;\;\vec{K}(\vec{f})=\vec{k}}\Theta(\vec{f})\langle\vec{0}|\vec{V}^{\cdot{\vec{k}}\dag}\left(i\vec{T}\right)^{\cdot\vec{N}(\vec{f})}|0\rangle=0\label{eq:fixing_equations}\\
	\sum_{f;\; \vec{K}(\vec{f})=\vec{k}}\Theta(\vec{f})\langle\vec{s}\neq 0|\vec{V}^{\cdot{\vec{k}}\dag}\left(i\vec{T}\right)^{\cdot\vec{N}(\vec{f})}|0\rangle=0\label{eq:backaction_equations}.
	\end{align}
	Eqs.~\ref{eq:backaction_equations} contain what we call the back-action terms.
	These are undesirable; if one fixes the $\theta_i^{(\vec{k})}$ values one at a time, then any non-zero term appearing in Eqs.~\ref{eq:backaction_equations} will need to be cancelled out by fixing some other $\theta^j_{\vec{k}'}$ at a later point.
	However, these terms may be avoided for a large class of parent ansatzes:
	\begin{dfn}
		\label{dfn:generating}
	A Pauli-type ansatz $(\prod_ie^{iT_i\theta_i},|\vec{0}\>)$ is \textbf{generating} if, for all computational basis states $|\vec{s}\>\neq|\vec{0}\>$, there exist generators $T_{\vec{s},a}$ for $a=0,1$ such that $iT_{\vec{s},a}|\vec{0}\>=i^a|\vec{s}\>$.
	\end{dfn}
	Note that a generating ansatz requires at least sufficient parameters to span the entire Hilbert space, however it remains unclear whether a generating ansatz does span the entire Hilbert space.
	Instead, we are interested in generating ansatzes here as they avoid undesired back-action
	\begin{lem}
	Given a generating Pauli-type variational ansatz $(\prod_{\vec{s},a}e^{iT_{\vec{s},a}\theta_{\vec{s},a}},|\vec{0}\>)$, one may solve Eqs.~\ref{eq:fixing_equations} by fixing $\theta_{\vec{s},a}^{(\vec{k})}=0$ unless $\vec{s}=\vec{s}(\vec{k})$ and $a=a(\vec{k}):=\Gamma(\vec{k})\;\mod 2$.
	This solution further prevents undesired back-action by making Eqs.~\ref{eq:backaction_equations} zero term-wise.\label{lem:only_diagrams}
	\end{lem}
	\emph{Proof ---} Eq.~\eqref{eq:fixing_equations} may be rewritten as
	\begin{align}
	&\sum_ai^{a-\Gamma(\vec{k})}\theta_{\vec{s}(\vec{k}),a}^{(\vec{k})}\nonumber\\ &\hspace{0.5cm}= C_{\vec{k}} - \sum_{\substack{f;\vec{K}(\vec{f})=\vec{k}\\|\vec{N}(\vec{f})|>1}}\Theta(\vec{f})\langle 0|\vec{V}^{\cdot\vec{k}\dag}\left(i\vec{T}\right)^{\cdot\vec{N}(\vec{f})}|0\rangle.\label{eq:fixing_rearranged}
	\end{align}
	We then use this equation to fix the left-hand side, being an equation of free $\Theta(\vec{f})$ terms.
	If this is done in ascending order in $|\vec{k}|$, one can check that all $\Theta(\vec{f})$ terms on the right-hand side at each $\vec{k}$ will have been fixed previously, implying that this fixing is well-defined.
	Then, one notes that
	\begin{equation}
	\langle 0|\vec{V}^{\cdot m\vec{k}\dag}\left(i\vec{T}\right)^{\cdot\vec{N}(md^{\vec{k},i})}|0\rangle=\langle 0|\vec{V}^{\cdot \vec{k}\dag}\left(i\vec{T}\right)^{\cdot\vec{N}(d^{\vec{k},i})}|0\rangle,
	\end{equation}
	for any odd $m$, which implies that contributions from linear combinations of the fixed components will never appear in Eq.~\eqref{eq:backaction_equations}.

	The above implies that the (strictly real) term $C_{\vec{k}}$ from each perturbative diagram contributes only to $\theta^{(k)}_{\vec{s}(\vec{k}),a(\vec{k})}$.
	Then, by definition, we have
	\begin{equation}
	i T_{\vec{s}(\vec{k}),a(\vec{k})}|\vec{0}\rangle=\pm\vec{V}^{\cdot\vec{k}}|\vec{0}\rangle,
	\end{equation}
	and as Pauli operators are either entirely real or entirely imaginary, this extends to any computational basis state $|\vec{s}'\rangle$
	\begin{equation}
	i T_{\vec{s}(\vec{k}),a(\vec{k})}|\vec{s}'\rangle=\pm\vec{V}^{\cdot\vec{k}}|\vec{s}'\rangle.
	\end{equation}
	This implies that for any function $\vec{f}$ such that $f_{s,a}(\vec{k})=0$, unless $\vec{s}=\vec{s}(\vec{k}),a=a(\vec{k})$ we have
	\begin{align}
	\left(i\vec{T}\right)^{\cdot\vec{N}(\vec{f})}|\vec{0}\rangle&=\pm\prod_{\vec{k}}\left(iT_{\vec{s}(\vec{k}),a(\vec{k})}\right)^{f_{\vec{s}(\vec{k}),a(\vec{k})}(\vec{k})}|\vec{0}\rangle\nonumber\\&=\pm\prod_{\vec{k}}\vec{V}^{\cdot f_{\vec{s}(\vec{k}),a(\vec{k})}(\vec{k})\vec{k}}|\vec{0}\rangle=\pm\vec{V}^{\vec{K}(\vec{f})}|\vec{0}\rangle,
	\end{align}
	and so the right-hand side of Eq.~\eqref{eq:fixing_rearranged} is real, and $\theta^{(\vec{k})}_{\vec{s}(\vec{k}),1-a(\vec{k})}=0$, by induction in $|\vec{k}|$.
	\qed

	For carefully-chosen Pauli-type ansatzes, one may further cancel contributions from disconnected diagrams.
	This yields our formal definition of what it means for such an ansatz to be `size-extensive' (as discussed in Sec.~\ref{sec:size_extensive})
	\begin{dfn}\label{dfn:size_extensive}
	We say that a Pauli-type ansatz $U(\vec{\theta})$ is size-extensive with respect to a perturbation $JV$ (Eq.~\eqref{eq:PT_ham}) if, in a solution to Eqs.~\ref{eq:fixing_equations}, $\theta^{(\vec{k})}_{\vec{s}(\vec{k}),a}=0$ if $\vec{k}=\vec{k}_A+\vec{k}_B$ is disconnected (Def.~\ref{dfn:disconnected_contributions}).
	\end{dfn}
	A Pauli-type ansatz satisfying this definition will satisfy Def.~\ref{dfn:size_extensive_informal} whenever the perturbative expansion above converges.
	To see this, note that when the perturbative expansion converges the solution to Eqs.~\ref{eq:fixing_equations} will provide the ground state exactly.
	Then, consider a Hamiltonian that does not couple two systems $S_i$ and $S_j$, and a term $T_{\vec{s},a}$ in our ansatz that does couple $S_i$ and $S_j$.
	One can see that whenever $\vec{s}=\vec{s}(\vec{k}),a=a(\vec{k})$ for some $\vec{k}$ that $\vec{k}$ will be disconnected, and so $\theta_{\vec{s},a}=0$ at all orders of $k$ by Def.~\ref{dfn:size_extensive}.

    We now have the machinery to present a condition for our ansatz to be size-extensive that just relates the ansatz terms $T_i$ to the perturbation terms $V_i$.
	\begin{dfn}
		\label{dfn:matched}
	A generating Pauli-type ansatz is \textbf{matched} to a perturbation $JV$ if
	\begin{align}
	\langle 0|\vec{V}^{\cdot\vec{k}\dag}\left(i\vec{T}\right)^{\cdot{\vec{N}(\vec{f})}}|0\rangle&\langle 0|\vec{V}^{\cdot\vec{k}'\dag}\left(i\vec{T}\right)^{\cdot\vec{N}(\vec{f}')}|0\rangle\nonumber\\&=\langle 0|\vec{V}^{\cdot(\vec{k}+\vec{k}')\dag}\left(i\vec{T}\right)^{\cdot\vec{N}(\vec{f}+\vec{f}')}|0\rangle,
	\end{align}
	whenever $(\vec{k},f)$ and $(\vec{k}',\vec{f}')$ act non-trivially on disconnected parts of the system.
	\end{dfn}

	\begin{exm}
	Any generating variational ansatz $(\prod_{\vec{s},a}e^{T_{\vec{s},a}\theta_{\vec{s},a}},|\vec{0}\>)$ for which the generators $T_{\vec{s},a}$ are compact (i.e. they only act nontrivially on qubit $j$ if $s_j=1$), is matched. In particular, QCA (Example \ref{exm:QCA}) is both generating and matched.
	\end{exm}
	
	\begin{thm}
	A perturbative hierarchy constructed from a Pauli-type ansatz via Eqs.~\ref{eq:fixing_equations}, that is matched to a perturbation $JV$, is size-extensive.\label{thm:no_disconnected_diagrams}
	\end{thm}
	\emph{Proof ---} By Lemma~\ref{lem:disconnected_contributions}, we have that $C_{\vec{k}}=C_{\vec{k}_A}C_{\vec{k}_B}$. Inserting Eq.~\eqref{eq:fixing_equations}, we find
	\begin{align}
	C_{\vec{k}}&=\sum_{f_A,\vec{K}(\vec{f}_A)=\vec{k}_A}\sum_{f_B,\vec{K}(\vec{f}_B)=\vec{k}_B}\Theta(\vec{f}_A)\Theta(\vec{f}_B)\nonumber\\&\times\langle\vec{0}|\vec{V}^{\cdot\vec{k}_A\dag}\left(i\vec{T}\right)^{\cdot\vec{N}(\vec{f}_A)}|\vec{0}\rangle\langle\vec{0}|\vec{V}^{\cdot\vec{k}_B\dag}\left(i\vec{T}\right)^{\cdot\vec{N}(\vec{f}_B)}|\vec{0}\rangle.
	\end{align}
	As disconnected parts of $\vec{k}$, either $k_{A,i}=0$ or $k_{B,i}=0$ for any $i$, implying $f_{A}(\vec{k}')=\vec{0}$ or $f_{B}(\vec{k}')=\vec{0}$ for all $\vec{k}'$ in the above sum.
	From this we may write
	\begin{align}
	\Theta(\vec{f}_A+\vec{f}_B)&=\prod_{\vec{k}',i}\frac{\left[\theta_i^{(\vec{k})}\right]^{f_A(\vec{k}')+\vec{f}_B(\vec{k}')}}{(\vec{f}_A(\vec{k}')+\vec{f}_B(\vec{k}'))!}\nonumber\\&=\Theta(\vec{f}_A)\Theta(\vec{f}_B).
	\end{align}
	Combining this with the definition of a matched ansatz obtains
	\begin{equation}
	C_{\vec{k}}=\sum_{\substack{f_A,\vec{K}(\vec{f}_A)=\vec{k}_A\\f_B,\vec{K}(\vec{f}_B)=\vec{k}_B}}\Theta(\vec{f}_A+\vec{f}_B)\langle\vec{0}|\vec{V}^{\vec{k}\dag}\left(i\vec{T}\right)^{\vec{N}(\vec{f}_A+\vec{f}_B)}|\vec{0}\rangle.\label{eq:combiningckackb}
	\end{equation}
	It remains to check that all $f:\NN^{\numcouplings}\rightarrow\NN^{\numparams}$ with $\vec{K}(\vec{f})=\vec{k}$, $|\vec{N}(\vec{f})|>1$, and $\Theta(\vec{f})\neq 0$ take the form $\vec{f}=\vec{f}_A+\vec{f}_B$ with $\vec{K}(\vec{f}_A)=\vec{k}_A$ and $\vec{K}(\vec{f}_B)=\vec{k}_B$, in which case the right-hand side of Eq.~\eqref{eq:fixing_rearranged} cancels, giving the required result. 
	This may be seen by induction in $|\vec{K}(\vec{f})|$.
	Clearly it is true for $|\vec{K}(\vec{f})|=1$.
	Then, fix $f$ with $|\vec{K}(\vec{f})|>1$, and define $f_A(\vec{k}')=f(\vec{k}')$ if $\vec{k}'_i\vec{k}_{B,i}=0$ for all $i$ and $f_A(\vec{k}')=0$ otherwise, and similarly for $f_B(\vec{k}')$, and define $f_{AB}=f-f_A-f_B$.
	One has that $\Theta(\vec{f})=\Theta(\vec{f}_A)\Theta(\vec{f}_B)\Theta(\vec{f}_{AB})$, but if $f_{AB}\neq 0$, it is a product of $\theta_{\vec{s}(\vec{k}_{AB}),a}^{(\vec{k}_{AB})}$ for disconnected $\vec{k}_{AB}$ with $|\vec{k}_{AB}|<K$, and thus $\Theta(\vec{f}_{AB})=0$.\qed
	
	This result can be seen as the digital quantum cousin of the linked-cluster theorem \cite{brueckner55}.

	\subsection{The perturbative construction \label{sec:PT_ansatz_hierarchy}}

	Following the above, we can construct a hierarchy of the $T_{\vec{s},a}$ by estimating the corresponding value of $\theta_{\vec{s},a}$ and placing them in order.
	We do not need to know the precise values of $\theta_{\vec{s},a}$, as these will be optimized as part of the VQE.
	Instead we plan to estimate only the largest contributions to each $\theta_{\vec{s},a}$.
	Under the assumption that $J_i~J\ll h_n$ for all interaction terms $i$ and all qubits $n$, we expect the largest contributions to come from those (connected) $C_{\vec{k}}$ with smallest possible $|\vec{k}|$.
	This may be read off immediately from the perturbative diagrams themselves
	\begin{dfn}
		A connected perturbative diagram $D$ for a vector $\vec{k}$ is a sub-leading diagram to a diagram $D'$ for a vector $\vec{k}'$ if:
		\begin{itemize}
			\item $D$ and $D'$ have identically coloured vertices (implying $\vec{s}(\vec{k})=\vec{s}(\vec{k}')$).
			\item $D$ and $D'$ have the same number of red edges modulo $2$ (implying $a(\vec{k})=a(\vec{k}')$).
			\item $D'$ has fewer interaction vertices than $D$ (implying $|\vec{k}|<|\vec{k}'|$).
		\end{itemize}
		A diagram $D$ is leading if it is not a sub-leading diagram to any $D'$.
	\end{dfn}
	Note that multiple leading diagrams may exist for a single parameter $\theta_{\vec{k}}^a$.

	We now wish to construct a perturbative hierarchy by drawing all leading diagrams with $|\vec{k}|<K$ interaction vertices (for some sufficiently large $K$), and then ordering corresponding $T_{\vec{s}}^{a}$ by the leading-order contributions to $\theta_{\vec{s},a}$ we obtain via Eq.~\eqref{eq:fixing_rearranged}.
	However, this calculation requires the normalized coefficients $C_{\vec{k}}$, which in turn require computing the perturbative series for the normalization constant $\Nn$.
	To avoid this cumbersome normalization procedure and also to simplify Eq.~\eqref{eq:fixing_rearranged}, we suggest to approximate $\theta^{(\vec{k})}_{\vec{s}(\vec{k}),a(\vec{k})}$ by 
	\begin{equation}
	\tilde{\theta}_{\vec{s},a}=\sum_{\substack{\mathrm{leading}\;\vec{k},\\\;\vec{s}(\vec{k})=\vec{s},\;a(\vec{k})=a}}\tilde{\theta}_{\vec{s}(\vec{k}),\vec{a}(\vec{k})}^{(\vec{k})}
	\end{equation}
	where we define
	\begin{align}
	&\hspace{-1cm}\sum_ai^{a-\Gamma(\vec{k})}\tilde{\theta}_{\vec{s}(\vec{k}),a}^{(\vec{k})} = \tilde{C}_{\vec{k}} - \sum_{\substack{f;\vec{K}(\vec{f})=\vec{k}\\|\vec{N}(\vec{f})|>1}}\tilde{\Theta}(\vec{f})\label{eq:fixing_tilde},\\
	\tilde{\Theta}(\vec{f}) &= \prod_{\vec{k},i}\frac{\left[\tilde{\theta}_i^{(\vec{k})}\right]^{f_i(\vec{k})}}{f_i(\vec{k})!}.
	\end{align}
	We expect that typically $\tilde{\theta}_{\vec{s},a}<\tilde{\theta}_{\vec{r},b}\leftrightarrow \theta_{\vec{s},a}<\theta_{\vec{r},b}$, which implies that this approximation should preserve the perturbative hierarchy.
	
	We now have all the machinery required to define our perturbative hierarchy.
	\begin{dfn}\label{dfn:perturbative_hierarchy}
	Let $\{T_{\vec{s},a}\}$ be the generators for a matched, generating variational ansatz for a Hamiltonian $H=H_0+\vec{J}\cdot\vec{V}$.
	The perturbative hierachy on $\{T_{\vec{s}},a\}$ is defined by the total order
	\begin{equation}
	T_{\vec{s},a} < T_{\vec{r},b} \; \mathrm{if} \; \tilde{\theta}_{\vec{s},a} < \tilde{\theta}_{\vec{r},b},
	\end{equation}
	and if $\tilde{\theta}_{\vec{s},a}=\tilde{\theta}_{\vec{r},b}$, we choose the ordering of $T_{\vec{s},a}$ and $T_{\vec{r},b}$ at random.
	\end{dfn}

	The explicit calculation of the $\tilde{\theta}_{\vec{s},a}$ variables is quite time consuming.
	As a shortcut, we note that $\tilde{\theta}^{(\vec{k})}_{\vec{s},a}$ scales as $\vec{J}^{\cdot\vec{k}}$, which, when $J_i\ll1$ typically dominates any combinatorial terms.
	To formalize this, let us define
	\begin{equation}
	J_{\vec{s},a}=\sum_{\substack{\mathrm{leading}\;\vec{k},\\\;\vec{s}(\vec{k})=\vec{s},\;a(\vec{k})=a}}\vec{J}^{\cdot\vec{k}},
	\end{equation}
	and we suggest to save on calculation by assuming $\theta_{\vec{s},a}<\theta_{\vec{r},b}$ when $J_{\vec{s},a}<J_{\vec{r},b}$.
	
	\section{Application: transverse-field Ising model\label{chapt:TFIM}}

	In this section, we demonstrate the construction of a variational hierarchy and study the resulting VQE performance on a target system.
	As a simple target example, we take the 1-dimensional transverse-field Ising model (TFIM):
	\begin{equation}
		H_{TFIM}=-\sum^{\numqubits}_i hZ_i+\sum^{\numqubits-1}_{i=1} JX_iX_{i+1}.
		\label{eq:TFIM}
	\end{equation}
	This system is a well-known prototype for condensed matter systems, being a non-interacting set of spins at $J=0$, an Ising chain at $h=0$, and demonstrating a quantum phase transition at $h=J$.
	For our example, we consider the $h\gg J>0$ regime, and construct a perturbative hierarchy around $J=0$, using the QCA as a parent ansatz.
	The noninteracting ground state may be immediately identified as the computational basis state $|\vec{0}\>$ with energy $-h\numqubits$, which we use as the starting state of our ansatz.
	Non-interacting excited states $|\vec{s}\>$ have energy $(2|\vec{s}|-\numqubits)h$.	
	\begin{figure}
	\includegraphics[width=\columnwidth]{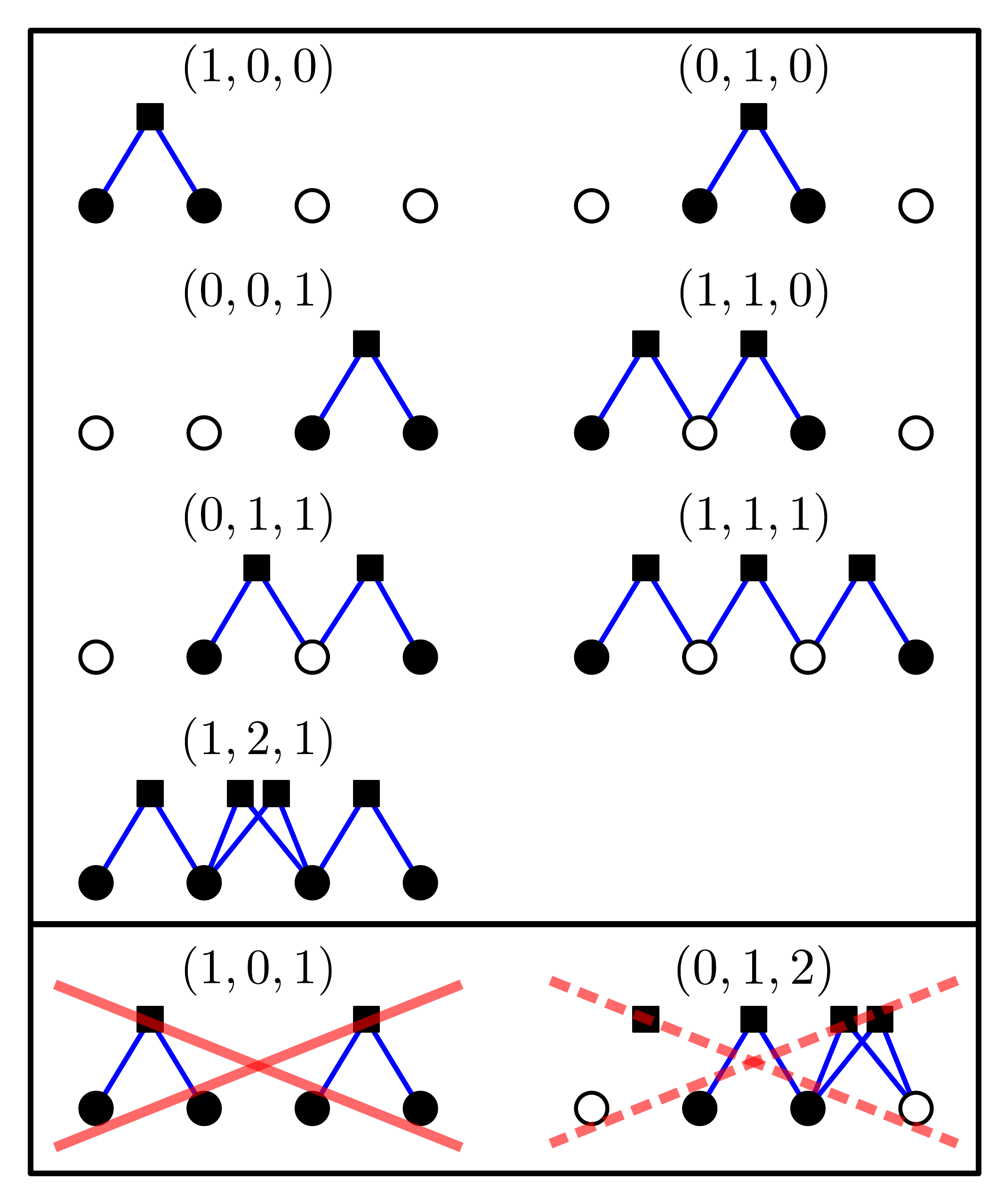}
	\caption{\label{fig:TFIM_foursites_diagrams} (top) The seven lowest-order connected diagrams for a four-site transverse-field Ising model, labeled by the $\vec{k}$ used in the text. (bottom) Examples of diagrams that do not need to be considered when constructing the perturbative hierarchy - (bottom left) a disconnected diagram that explicitly does not contribute to the hierarchy, and (bottom right) a diagram which will contribute to the same parameter in the hierarchy as a previous term ($\vec{k}=(1,1,0)$), but to lower order.}
	\end{figure}
	\subsection{Example perturbative construction on four sites\label{sec:gate_hierarchies_second_order}}

	To demonstrate the application of the methods developed in Sec.~\ref{sec:hierarchies} in detail, we now construct the full perturbative hierarchy on a small chain ($\numqubits=4$).
	This system has three perturbation terms, which we label $\hat{V}_i=X_iX_{i+1}$ for $i=1,2,3$.
	These perturbations preserve the antiunitary complex conjugation symmetry $\Kk$, and the unitary global parity symmetry $Z_1Z_2Z_3Z_4$.
	This reduces the required variational manifold dimension from $2^5-2=30$ to $2^3-1=7$ (both symmetries halve the Hilbert space dimension, but complex conjugation makes the phase equivalence redundant).
	In the QCA, this corresponds to removing all imaginary rotations (of the form $e^{i\theta X...X}$), and all generators with an odd number of non-trivial terms.
	This removal will be automatic in the perturbative construction, as removed terms will never appear in the hierarchy, so we need only note the symmetries in case we `run out' of terms to add to the variational ansatz~\footnote{Note that this is not always the case: if one must satisfy a symmetry by fixing parameters, both terms will appear in the hierarchy and the fixing must be done after the hierarchy is constructed.}.
	The remaining generators are then
	\begin{align*}
	T_1&=X_1Y_2,\hspace{0.5cm} T_2=X_2Y_3,\hspace{0,5cm} T_3=X_3Y_4,\\
	T_4&=X_1Y_3,\hspace{0.5cm} T_5=X_2Y_4,\hspace{0.5cm} T_6=X_1Y_4,\\
	T_7&=X_1X_2X_3Y_4.
	\end{align*}
	For convenience in this small system, we will drop the stabilizer notation of Sec.\ref{sec:stablizer_ansatze}, and write the QCA as $\prod_{j=1}^7\exp(i\theta_jT_j)$.
	(For example, in the notation of Sec.~\ref{sec:stablizer_ansatze} we would have written $\theta_6$ as $\theta^4_{XII,1}$.)

	To construct the perturbative hierachy, we proceed by drawing all lowest-order diagrams, and calculating the corresponding $\tilde{C}_{\vec{k}}$ contributions.
	In Fig.~\ref{fig:TFIM_foursites_diagrams}, we list the seven lowest-order connected diagrams in the system.
	This gives us the following:
	\begin{enumerate}
	\item 3 contributions at order $J$ (to $T_1,T_2$, and $T_3$).
	\item 2 contributions at order $J^2$ (to $T_4$ and $T_5$).
	\item 1 contribution at order $J^3$ (to $T_6$).
	\item 1 contribution at order $J^4$ (to $T_7$).
	\end{enumerate}
	This may then be used as an initial guess for the ordering in the perturbative hierarchy.
	Importantly, although $\vec{k}=(1,0,1)$ is an order-$J^2$ term satisfying $\langle 0|\vec{V}^{\vec{k}}T_7|0\rangle\neq 0$, the corresponding diagram is disconnected (Fig.~\ref{fig:TFIM_foursites_diagrams}, bottom-left).
	This implies that its contribution to $\theta_7$ will be cancelled out by the contributions of $(1,0,0)$ and $(0,0,1)$ (Theorem \ref{thm:no_disconnected_diagrams}), and the diagram need not be considered in our construction, as we will confirm shortly.
	We further note that higher-order diagrams exist, e.g. that corresponding to $\vec{k}=(0,1,2)$ (Fig.~\ref{fig:TFIM_foursites_diagrams}, bottom-right).
	Although these have non-zero contribution to the actual value of the variational angles (in this case $\theta_2$), as this contribution is at a higher-order of $J$ we expect it to not affect the order of the hierarchy.

	We now check the above ordering of the perturbative hierarchy by explicit calculation of the lowest-order contributions to $\tilde{\theta}_j$.
	Applying Eq.~\eqref{eq:tilde_Ck} recursively, the lowest-order connected contributions can be found to be (noting $S_{\vec{k},\vec{k}'}=1$ as all $V_i$ commute),
	\begin{align*}
	\tilde{C}_{(1,0,0)}&=\tilde{C}_{(0,1,0)}=\tilde{C}_{(0,0,1)}=\frac{-1}{4h}\\
	\tilde{C}_{(1,1,0)}&=\frac{-1}{4h}[\tilde{C}_{(1,0,0)}+\tilde{C}_{(0,1,0)}]=\frac{1}{8h^2}\left[=\tilde{C}_{(0,1,1)}\right]\\
	\tilde{C}_{(1,1,1)}&=\frac{-1}{4h}[\tilde{C}_{(1,1,0)}+\tilde{C}_{(1,0,1)}+\tilde{C}_{(0,1,1)}]=-\frac{5}{64h^3}\\
	\tilde{C}_{(1,2,1)}&=\frac{-1}{8h}[\tilde{C}_{(0,2,1)}+\tilde{C}_{(1,2,0)}+\tilde{C}_{(1,1,1)}\\
	&\hspace{1cm}-\tilde{C}_{(0,1,0)}\tilde{C}_{(1,0,1)}]=\frac{3}{256h^4}.
	\end{align*}
	One may then calculate in turn the lowest-order approximation for the variational parameters via Eq.~\eqref{eq:fixing_tilde} (noting here that $\Gamma(\vec{k})=1$ for all $\vec{k}$ in this system).
	\begin{align*}
	\tilde{\theta}_1&=J\tilde{C}_{(1,0,0)}=\frac{-J}{4h}\left[=\tilde{\theta}_2=\tilde{\theta}_3\right]\\
	\tilde{\theta}_4&= J^2\tilde{C}_{(1,1,0)}-\tilde{\theta}_1\tilde{\theta}_2=\frac{J^2}{16h^2}\left[=\tilde{\theta}_5\right]\\
	\tilde{\theta}_6&= J^3\tilde{C}_{(1,1,1)}-\tilde{\theta}_1\tilde{\theta}_2\tilde{\theta}_3-\tilde{\theta}_1\tilde{\theta}_5-\tilde{\theta}_3\tilde{\theta}_4=-\frac{J^3}{32h^3}\\
	\tilde{\theta}_7&= J^4\tilde{C}_{(1,2,1)}-\frac{1}{2}\tilde{\theta}_1\tilde{\theta}_2^2\tilde{\theta}_3\\&\hspace{1cm}-\tilde{\theta}_4\tilde{\theta}_2\tilde{\theta}_3-\tilde{\theta}_1\tilde{\theta}_2\tilde{\theta}_5-\tilde{\theta}_4\tilde{\theta}_5=-\frac{J^4}{512h^4}.
	\end{align*}
	We see that ordering terms by $J_{\vec{s},a}$ reproduces the full perturbative hierarchy whenever $J<2h$.
	We also note that the order $J^2$ contribution to $\tilde{\theta}_7$ from $\vec{k}=(1,0,1)$ is cancelled (following Theorem \ref{thm:no_disconnected_diagrams}), as
	\begin{equation}
	J^2\tilde{C}_{(1,0,1)}=\frac{J^2}{16h^2}=\tilde{\theta}_1\tilde{\theta}_3.
	\end{equation}
	We also note that the magnitudes of $\tilde{\theta}_i$ are systematically smaller than the magnitudes of corresponding perturbative terms $\vec{J}^{\cdot\vec{k}}\tilde{C}_{\vec{k}}$. 
	This suggests that the back-action terms $\sum_{\substack{f;\vec{K}(\vec{f})=\vec{k};|\vec{N}(\vec{f})|>1}}\Theta(\vec{f})\langle 0|\vec{V}^{\cdot\vec{k}\dag}\left(i\vec{T}\right)^{\cdot\vec{N}(\vec{f})}|0\rangle$ in QCA may have a systematic positive effect on VQE convergence.
	
	\subsection{Low-order construction for a large chain \label{sec:TFIM_PT_and_ansatze}}

	\begin{figure}
	\includegraphics[width=\columnwidth]{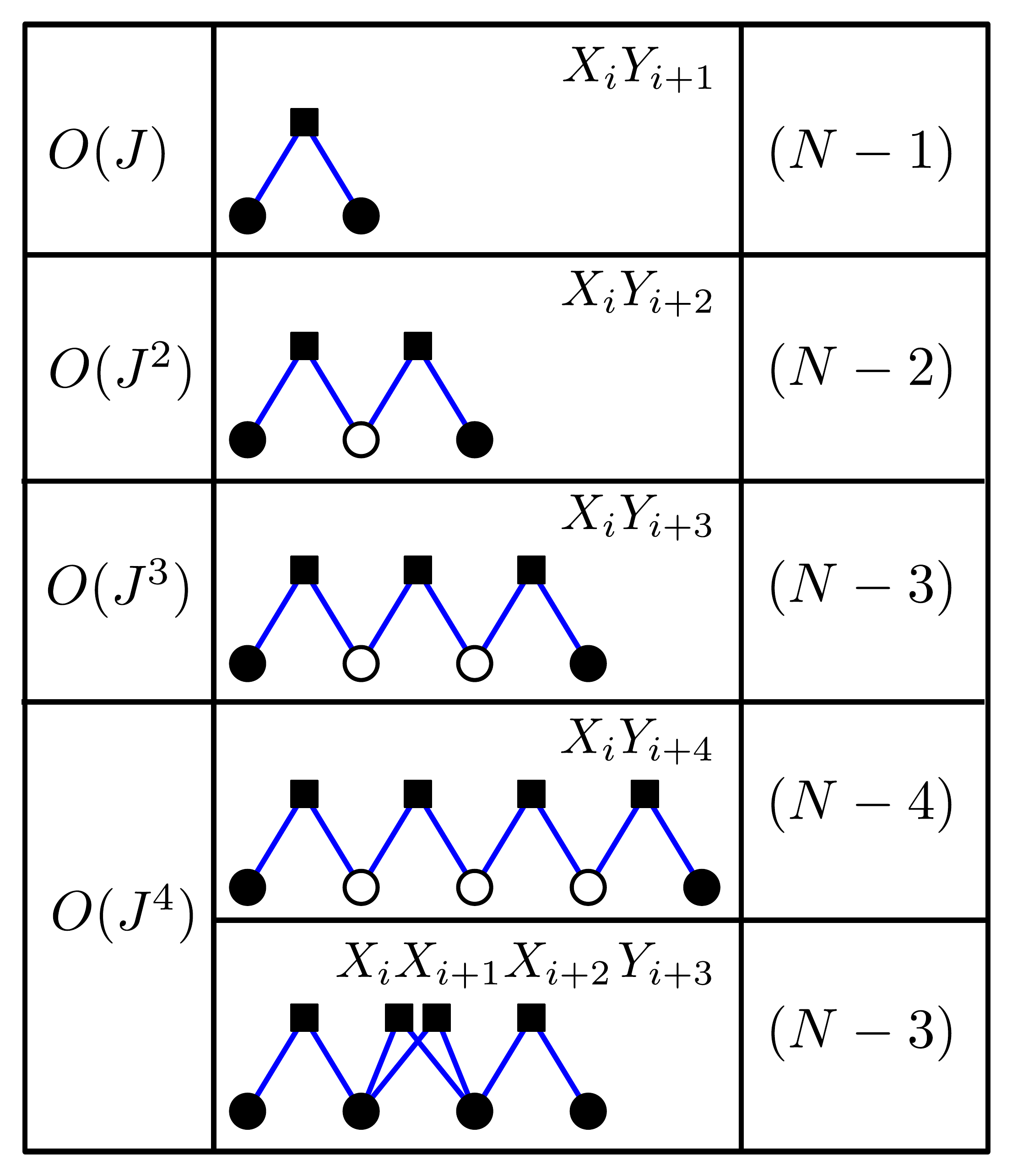}
	\caption{\label{fig:fourth_order_connected}The leading connected diagrams to fourth-order on the transverse-field Ising model. Each diagram should be repeated across the entire $\numqubits$-qubit chain - the total number of copies of each diagram that will appear is written in the right-hand column. Diagrams are labelled by the generator $T_{\vec{s},a}$ that they contribute to.}
	\end{figure}

	Following the analysis of the four-site example, we expect little to no deviation between parameters of the same order in a larger chain.
	Indeed, all first, second and third-order leading diagrams are identical up to translation along the chain (Fig.~\ref{fig:fourth_order_connected}).
	As the on-site and interaction strengths are uniform along the chain, this implies that the coefficients for all such diagrams are likewise equal (to lowest-order).
	At fourth-order, two separate types of diagrams exist.
	One corresponds to $\vec{k}=(1,2,1)$ in the four-site model, and gives the same parameter estimate ($\tilde{\theta}_{\vec{s},a}=\frac{-J^4}{512h^4}$), to the QCA generators of the form $\{Y_iX_{i+1}X_{i+2}X_{i+3}\}$
	The other was not present in the four-site model (as it requires $5$ qubits) - it contributes a parameter estimate of $\tilde{\theta}_{\vec{s},a}=\frac{J^4}{128h^4}$ to QCA generators of the form $\{Y_iX_{i+4}\}$, placing these generators earlier in the perturbative hierarchy. 
	The resulting ansatz thus needs only $5\numqubits-13$ generators to reproduce the ground state with errors of order $(J/h)^5$.
	To obtain this level of accuracy with a classical calculation, one would in theory need to sum over all $(\numqubits-1)^4$ combinations of individual perturbations.
	However, as clever grouping of terms (e.g. via tensor network contractions or similar) should reduce the time-cost of such a summation far below such numbers, this argument does not lead to an immediate guarantee of a quantum speedup for VQEs of this form.

	\subsection{Alternative hierarchies and circuit ordering}
	
	Although perturbation theory is a natural choice for developing variational hierarchies, it is not necessarily the only starting point.
	In the presence of strong interactions (where pertubation theory breaks down), other generator properties may provide better insight into how important they are at obtaining the ground state.
	In the following, we study the following natural constructions of a priority list, all of which use QCA as a parent ansatz:
	\begin{itemize}
		\item \emph{pertQCA:} The perturbative hierarchy from Def.~\ref{dfn:perturbative_hierarchy}, using QCA as the parent variational ansatz.
		\item \emph{revQCA:} The pertQCA hierarchy in reverse.
		\item \emph{2-locQCA:} A low-weight variant of pertQCA, obtained by only allowing $2$-local generators (those acting non-trivially on up to $2$ qubits). When more generators are desired than in the final priority list, we loop over it repeatedly.
 		\item \emph{locQCA:} A geometrically local variant of pertQCA, obtained by only allowing generators acting on nearest neighbour pairs of qubits (and again looping over the priority list if required). This is equivalent to allowing only the generators which are dictated by the first-order perturbation theory, allowing for a generalization to an arbitrary Hamiltonian.
	\end{itemize}
	
	We have so far not discussed the ordering of the units within the ansatz circuit.
	Two natural choices present themselves: taking the order in which the gates appear in the priority list, and taking the order in which the gates appear in the parent ansatz.
	However, this is only well-defined when the priority list is inherited from a parent ansatz without repetition.
	For the above hierarchies that require looping, we only study the former choice, and denote by an asterisk results where the latter ordering is used.
	
	\subsection{VQE performance \label{sec:VQE_performance}}

	We now test the performance of our variational hierarchies in different parameter regimes of the transverse-field Ising model on $\numqubits=8$ sites.
	(Code to perform this investigation can be found at \url{https://github.com/tarrlikh/QSA}.)
	We take as a performance metric the relative energy error
	\begin{equation}
	\epsilon := (E_{\mathrm{VQE}}-E_0)/E_0,\label{eq:rel_energy_error}
	\end{equation}
	where $E_{\mathrm{VQE}}$ is the energy of the converged VQE, and plot this as we increase the number $N_p$ of parameters in the hierarchy.
	The hierarchy gives a natural strategy to perform the optimization - at each $N_p$, the optimized values of the previous $N_p-1$ parameters are used as a starting guess for their new values (whilst the new parameter is initialized to $0$).
	This approach converges much faster than re-starting each new simulation at the original value, as found previously in~\cite{Romero18}.
	To focus on the performance of the ansatzes themselves, we do not include the effects of sampling noise or any experimental noise in our simulations.

	\begin{figure}
	\includegraphics[width=\columnwidth]{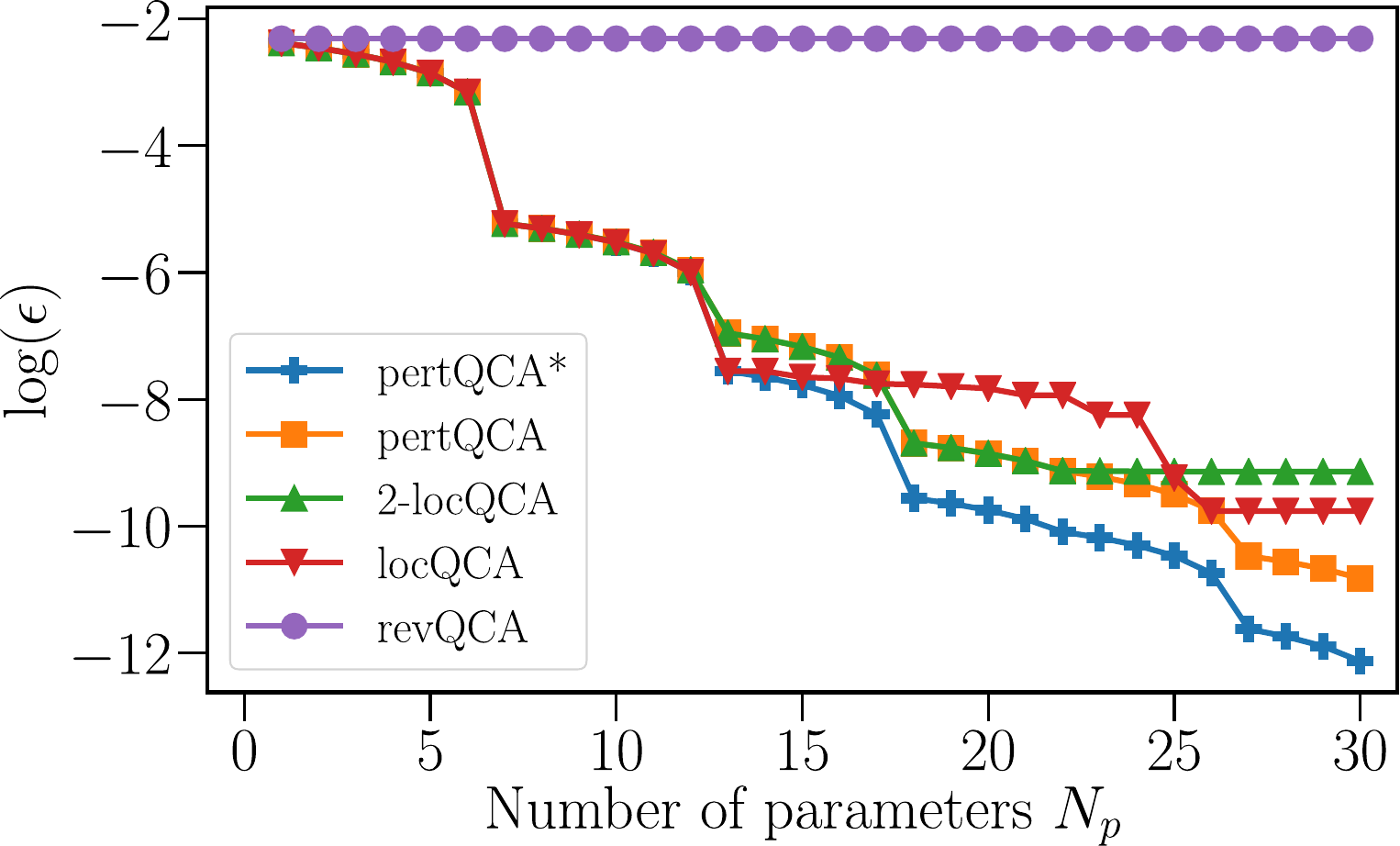}
	\caption{\label{fig:QCA_performance_weakcoupling}Log plot of the relative energy error $\epsilon$ (Eq.~\eqref{eq:rel_energy_error}) for different variational hierarchies, in a weakly-coupled transverse-field Ising model ($J/h=0.15$). Error is plotted as a function of the number of parameters used (or equivalently the number of generators taken from the hierarchy). Description of the different hierarchies is given in the text.}
	\end{figure}
	We first investigate the weak-coupling regime where perturbation theory holds ($J/h=0.15$).
	In Fig.~\ref{fig:QCA_performance_weakcoupling}, we plot the convergence of $\epsilon$ as the first $30$ terms from all studied hierarchies are added consecutively.
	At each subsequent point we reoptimize all parameters using the SLSQP algorithm, starting from the local minimum found at the previous point.
	We observe that all hierarchies achieve good convergence, with the exception of revQCA, and that both variants of pertQCA achieve over an order of magnitude improvement over other ansatzes after 30 terms are added.
	We further observe that re-ordering the gates to follow the parent ansatz (pertQCA*) is preferable, leading to another order of magnitude improvement.
	We are unsure of the precise reason for this improvement, but suggest it may be attributed to the relatively large area of the variational manifold inherited from the parent ansatz, that may be lost under re-ordering.
	The discontinuities in the plot for pertQCA, pertQCA*, and 2-locQCA correspond to the points where all gates up to a certain perturbation theory order have been included.
	This makes sense, as our theory predicts these points should correspond to the error decreasing from $O(J^n)$ to $O(J^{n+1})$.

	\begin{figure}
	\includegraphics[width=\columnwidth]{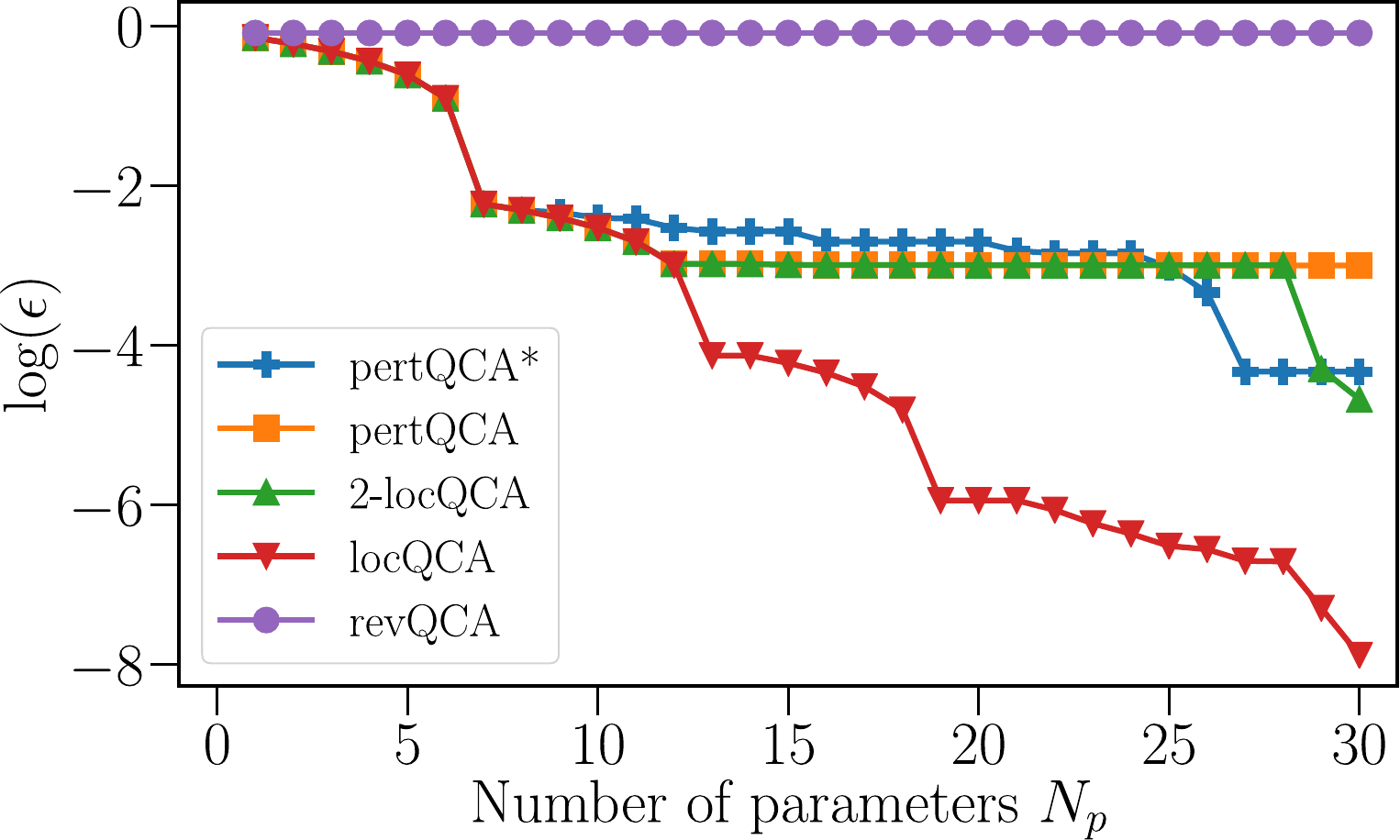}
	\caption{\label{fig:QCA_performance_strongcoupling}Similar convergence plot to Fig.~\ref{fig:QCA_performance_weakcoupling}, but in the strongly-coupled regime instead ($J/h=6$).}
	\end{figure}
	We next investigate VQE convergence in the strongly correlated regime ($J/h=6$).
	We observe that all hierarchies perform worse here than previously.
	We attribute this to the strongly-coupled ground state being further from the starting state than the weakly-coupled ground state.
	Note however, that one can obtain one of the two degenerate ground states at $h=0$ from $|\vec{0}\rangle$ as
	\begin{equation}
	|E_0(h=0)\rangle=\prod_ie^{i\frac{\pi}{4}X_iY_{i+1}}|\vec{0}\rangle,\label{eq:PT_trick}
	\end{equation}
	which is a rotation achievable after the first $\numqubits-1=7$ terms of all considered hierarchies.
	This suggests that in all cases, the first order of the hierarchy is used to prepare this state, from which later orders perturb.
	Then, as perturbation theory around the strongly correlated ground state is significantly different to the perturbation theory around the non-interacting ground state, the generators we have chosen may not be optimal for this perturbation.
	This also explains the good performance of locQCA over the other hierarchies: by repeating local operators it ensures that it will obtain the lower orders (in $h/J$) of the true ground state.

	\begin{figure}
	\includegraphics[width=\columnwidth]{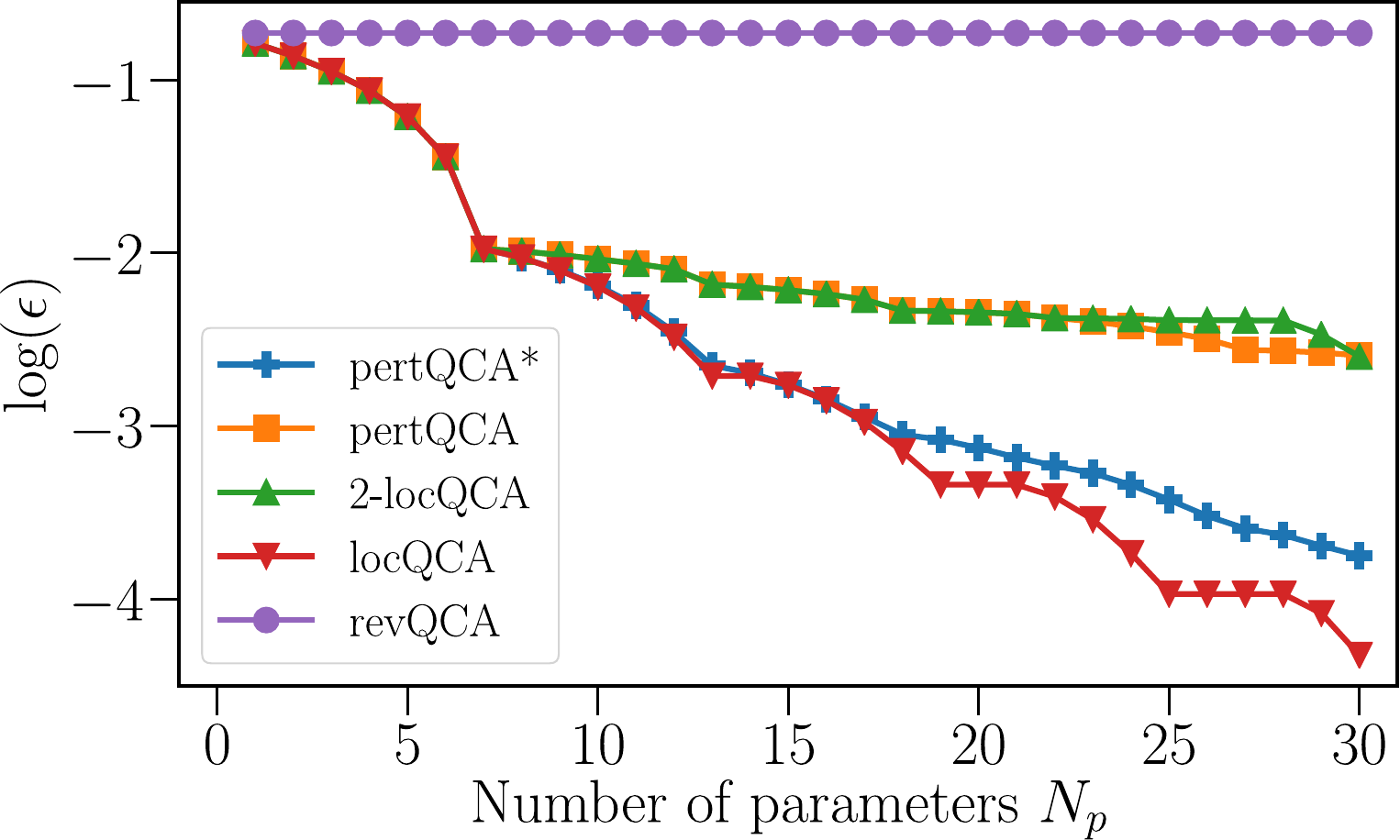}
	\caption{\label{fig:QCA_performance_critical}Similar convergence plot to Fig.~\ref{fig:QCA_performance_weakcoupling}, but in the critical regime instead ($J/h=1$).}
	\end{figure}
	We finally investigate the performance of our hierarchies in the critical regime ($J/h=1$), where a transition between the strongly-correlated and weakly-correlated phases occurs in the thermodynamic limit.
	We observe that the relative error obtained by all ansatzes is the worst here, and that locQCA and pertQCA* behave similarly, obtaining up to an order of magnitude improvement over 2-locQCA and pertQCA.
	This loss of accuracy is not surprising, as we do not have a relatively cheap way of accessing any states perturbatively coupled to the ground state in the same manner as Eq.~\eqref{eq:PT_trick}.

	\section{Conclusion}

	In this work, we have developed a diagrammatic framework for size-extensive variational quantum ansatzes, which avoids the use of Trotter-Suzuki approximation methods.
	We have described a large class of Pauli-generated product ansatzes demonstrably capable of spanning the entire Hilbert space with the minimum number of parameters necessary.
	We have demonstrated means by which one can compress ansatzes such as the above to a practical size, by a perturbative treatment of the target system, and by taking into account any symmetries that exist.
	To ensure the size-extensivity of the construction, we have stated and proven the digital quantum version of the linked-cluster theorem.
	We have tested variants of the resulting ansatzes on the transverse-field Ising model, finding that their performance in various regimes matches our expectations based on their means of construction.
	We observe that ansatzes that fully match the perturbation theory give a benefit in the weak coupling regime as expected.
	However, in the strong-coupling regime, focusing on the locality of the ansatz at the expense of perturbation theory considerations appears to be preferred.

    As is well known in the field, the performance of any VQE  ansatz is system dependent.
    Ansatzes that are derived from perturbative physical principles can be expected to perform best when perturbation theory converges well.
    By contrast, those founded on adiabatic principles (e.g. the variational Hamiltonian ansatz~\cite{Wecker14}) can be expected to perform best on systems with a large gap.
    As these two conditions are often correlated (e.g. a gap closing often corresponds to a phase transition and a breakdown of perturbation theory), a fair comparison of ansatzes based on these two principles (and with any other ansatzes) would require an extensive numerical study.
    This is an obvious target for future research.
	
	We have avoided in the above any discussion of a quantum speedup for the VQEs that we have constructed in this work.
	To the best of our knowledge this remains an open and difficult question to show for any class of VQEs.
	Informally, to demonstrate a quantum speedup, one requires to be able to obtain an estimate of the true ground state energy $E$ for an $\numqubits$-qubit system, within an error $\epsilon$, in time polynomial in $\numqubits$.
	This also needs to be achieved in a class of $\numqubits$-qubit systems for which no similar estimation is possible classically.
	The circuit length in a variational hierarchy grows polynomially in the number of parameters $N_p$, so it would be sufficient to show that the error $\epsilon(N_p,N)$ scales polynomially in $N_p$ and $\numqubits$.
	One also needs to consider the time cost of measuring the energy (which grows polynomially in $\numqubits$) and the time cost of optimization (which grows polynomially in $N_p$).
	Our results appear to show this behavior; we observe what appears to be exponential decay in $N_p$ for all three systems studied.
	(Note that the measurement and optimization requirements imply that the time cost to extract these energies from the device will still be at best polynomial.)
	However, $1$D spin chains such as the transverse-field Ising model are well accessible by classical methods and polynomial-time algorithms are known for any weakly-coupled 2-local spin system~\cite{BKL}, so we do not expect a quantum speedup in this case.
	Finding target systems for which a speedup may be demonstrable, and further optimizing hierarchy construction to show this, are obvious targets for future research.

	\begin{acknowledgements}
	We would like to thank X.~Bonet, B.~Terhal, V.~Cheianov, J.~Zaanen, V.~Lipinska, J.~Helsen, M.~Semenyakin, L.~Visscher and C.W.J.~Beenakker for support in this project. This work was funded by the Netherlands Organization for Scientific Research (NWO/OCW), an ERC Synergy Grant, and Shell Global Solutions BV.
	\end{acknowledgements}


\begin{thebibliography}{99}

\bibitem{Preskill18} J. Preskill, \href{https://doi.org/10.22331/q-2018-08-06-79}{\textit{Quantum Computing in the NISQ era and beyond}}, Quantum \textbf{2}, 79 (2018).

\bibitem{Litinski19} D. Litinski, \href{	https://doi.org/10.22331/q-2019-12-02-205}{\textit{Magic State Distillation: Not as Costly as You Think}}, Quantum \textbf{3}, 205 (2019).


\bibitem{Peruzzo14} A. Peruzzo, J. McClean, P. Shadbolt, M.-H. Yung, X.-Q. Zhou, P. J.Love, A. Aspuru-Guzik, and J. L. O’Brien, \href{https://doi.org/10.1038/ncomms5213}{\textit{A variational eigenvalue solver on a photonic quantum processor}}, Nat. Comm. \textbf{5}, 4213 (2014).

\bibitem{Mcclean16} J. R. McClean, J. Romero, R. Babbush, and A. Aspuru-Guzik, \href{https://doi.org/10.1088/1367-2630/18/2/023023}{\textit{The theory of variational hybrid quantum-classical algorithms}}, New J. Phys. \textbf{18}, 023023 (2016).

\bibitem{Romero18} J. Romero, R. Babbush, J. R. McClean, C. Hempel, P. J. Love, and A. Aspuru-Guzik, \href{https://doi.org/10.1088/2058-9565/aad3e4}{\textit{Strategies for quantum computing molecular energies using the unitary coupled cluster ansatz}} , Quantum Sci. Technol. \textbf{4}, 014008 (2018).

\bibitem{McClean18} J. McClean, S. Boixo, V. Smelyanskiy, R. Babbush, and H. Neven, \href{https://doi.org/10.1038/s41467-018-07090-4}{\textit{Barren plateaus in quantum neural network training landscapes}} , Nat. Comm. \textbf{9}, 4812 (2018).

\bibitem{DallaireDemers18} P.-L. Dallaire-Demers, J. Romero, L. Veis, S. Sim, and A. Aspuru-Guzik, \href{https://doi.org/10.1088/2058-9565/ab3951}{\textit{Low-depth circuit ansatz for preparing correlated fermionic states on a quantum computer}}, Quantum Sci. Technol. \textbf{4}, 045005 (2019).

\bibitem{Farhi14} E. Farhi, J. Goldstone, and S. Gutmann, \href{https://arxiv.org/abs/1411.4028}{\textit{A Quantum Approximate Optimization Algorithm}}, arXiv:1411.4028.

\bibitem{Lloyd18} S. Lloyd, \href{https://arxiv.org/abs/1812.11075}{\textit{Quantum approximate optimization is computationally universal}} , arXiv:1812.11075.

\bibitem{brueckner55} K. A. Brueckner, \href{https://doi.org/10.1103/PhysRev.100.36}{\textit{Many-Body Problem for Strongly Interacting Particles. II. Linked Cluster Expansion}}, Phys. Rev. \textbf{100}, 36 (1955).

\bibitem{trotter1959product} H. F. Trotter, \href{https://doi.org/10.1090/S0002-9939-1959-0108732-6}{\textit{On the Product of Semi-Groups of Operators}}, Proc. Am. Math. Soc. \textbf{10}, 545 (1959).

\bibitem{suzuki1991general} M. Suzuki, \href{https://doi.org/10.1063/1.529425}{\textit{General theory of fractal path integrals with applications to many‐body theories and statistical physics}}, J. Math. Phys. \textbf{32} (1991).

\bibitem{Whitfield11} J. D. Whitfield, J. Biamonte,  and A. Aspuru-Guzik, \href{https://doi.org/10.1080/00268976.2011.552441}{\textit{Simulation of electronic structure Hamiltonians using quantum computers}} , Mol. Phys. \textbf{109}, 735 (2011).

\bibitem{Kandala17} A. Kandala, A. Mezzacapo, K. Temme, M. Takita, M. Brink, J. M.Chow,  and J. M. Gambetta, \href{https://doi.org/10.1038/nature23879}{\textit{Hardware-efficient Variational Quantum Eigensolver for Small Molecules and Quantum Magnets}} , Nature \textbf{549}, 242 (2017).

\bibitem{Sagastizabal19} R. Sagastizabal, X. Bonet-Monroig, M. Singh, M. Rol, C. Bultink, X. Fu,  C.  Price, V. Ostroukh,  N.  Muthusubramanian,  A.  Bruno, M. Beekman, N. Haider, T. O’Brien, and L. DiCarlo, \href{https://doi.org/10.1103/PhysRevA.100.010302}{\textit{Error Mitigation by Symmetry Verification on a Variational Quantum Eigensolver}}, Phys. Rev. A \textbf{100}, 010302 (2019).

\bibitem{Guerreschi17}  G. Guerreschi and M. Smelyanskiy, \href{https://arxiv.org/abs/1701.01450}{\textit{Practical optimization for hybrid quantum-classical algorithms}}, arXiv:1701.01450.

\bibitem{higgott2018variational} O. Higgott, D. Wang, and S. Brierley, \href{https://doi.org/10.22331/q-2019-07-01-156}{\textit{Variational Quantum Computation of Excited States}}, Quantum \textbf{3}, 156 (2019).

\bibitem{endo2018variational} S. Endo, T. Jones, S. McArdle, X. Yuan, and S. Benjamin, \href{https://doi.org/10.1103/PhysRevA.99.062304}{\textit{Variational quantum algorithms for discovering Hamiltonian spectra}}, Phys. Rev. A \textbf{99}, 062304 (2019).

\bibitem{Nakanishi19} K. M. Nakanishi, K. Fujii, and S. Todo, \href{https://doi.org/10.1103/PhysRevResearch.2.043158}{\textit{Sequential minimal optimization for quantum-classical hybrid algorithms}}, Phys. Rev. Research \textbf{2}, 043158 (2020).

\bibitem{Gottesman97} D. Gottesman, \href{https://arxiv.org/abs/quant-ph/9705052}{\textit{Stabilizer Codes and Quantum Error Correction}}, PhD Dissertation, California Institute of Technology (1997).

\bibitem{Gard20} B.T. Gard, L. Zhu, G.S. Barron, N.J. Mayhall, S.E. Economou, and E. Barnes, \href{https://doi.org/10.1038/s41534-019-0240-1}{\textit{Efficient symmetry-preserving state preparation circuits for the variational quantum eigensolver algorithm}}, NPJ Quantum Inf. \textbf{6}, 10 (2020).

\bibitem{KT} J. Kirkwood and L. Thomas, \href{https://doi.org/10.1007/BF01211959}{\textit{Expansions and phase transitions for the ground state of quantum Ising lattice systems}}, Commun. Math. Phys. \textbf{88}, 569 (1983).

\bibitem{BKL} S. Bravyi, D. DiVincenzo,  and D. Loss, \href{https://doi.org/10.1007/s00220-008-0574-6}{\textit{Polynomial-time algorithm for simulation of weakly interacting quantum spin systems}} , Commun. Math. Phys. \textbf{284}, 481 (2008).

\bibitem{Wecker14} D. Wecker, M. B. Hastings, and M. Troyer, \href{https://doi.org/10.1103/PhysRevA.92.042303}{\textit{Progress towards practical quantum variational algorithms}}, Phys. Rev. A \textbf{92}, 042303 (2015).

\end{thebibliography}

	\pagebreak
	
	\appendix

	\section{Background}\label{app:prelim}
	
	\begin{dfn}
		The state of an $\numqubits$-qubit quantum register is represented by a norm-$1$ vector in the Hilbert space $\Hh=\mathbb{C}^{2^\numqubits}$, under the association $|\psi\>\in\Hh\equiv e^{i\phi}|\psi\>$ for $\phi\in\RR$.
	\end{dfn}
	\begin{dfn}
		The Pauli basis on $\numqubits$ qubits is defined as $\PP^\numqubits:=\{I,X,Y,Z\}^{\otimes \numqubits}$, where $I,X,Y,Z$ are the $2\times 2$ matrices on $\mathbb{C}^2$:
		\begin{align}
			I&=\left(\begin{array}{cc}1&0\\0&1\end{array}\right),X=\left(\begin{array}{cc}0&1\\1&0\end{array}\right),\nonumber\\ Y&=\left(\begin{array}{cc}0&-1i\\1i&0\end{array}\right),Z=\left(\begin{array}{cc}1&0\\0&-1\end{array}\right),
		\end{align}
		and $\otimes$ is the Kronecker tensor product.
	\end{dfn}
	$\PP^{\numqubits}$ has the following nice properties:
	\begin{enumerate}
		\item $P^2=1$ for all $P\in\PP^{\numqubits}$.
		\item For $P,Q\in\PP^{\numqubits}$, either $[P,Q]:=PQ-QP=0$, or $\{P,Q\}:=PQ+QP=0$, and $P$ commutes with precisely half of $\PP^{\numqubits}$.
		\item $P\in\PP^{\numqubits}\neq 1$ has only two eigenvalues, $\pm 1$, and the dimension of the corresponding eigenspaces is precisely $2^{\numqubits-1}$ (i.e. each $P$ divids $\mathbb{C}^{2^{\numqubits}}$ in two).
		\item This division by two may be further continued - given $P,Q\neq 1$ such that $[P,Q]=0$, $P$ and $Q$ divide the Hilbert space into $4$ eigenspaces (labeled by combinations of their eigenvalues).
		\item To generalize, one can form a $[\numqubits,k]$ \textbf{stabilizer group} $\Ss$, generated by $k$ Hermitian, commuting, non-generating elements of $\PP^{\numqubits}$ (up to a complex phase); this diagonalizes $\mathbb{C}^{2^{\numqubits}}$ into $2^k$ unique eigensectors of dimension $2^{\numqubits-k}$. When $\numqubits=k$, these sectors contain single eigenstates, which we call \textbf{stabilizer states}~\cite{Gottesman97}.
		\item Given such a stabilizer state $|\psi\>$ and Hermitian $P\in\PP^{\numqubits}$, either $P|\psi\>=\pm|\psi\>$ or $\<\psi|P|\psi\>=0$.
	\end{enumerate}
	The Pauli basis is a basis for the set of $2^{\numqubits}\times 2^{\numqubits}$ complex-valued matrices (hence the name); it is also a basis for the set of Hermitian matrices if one chooses real coefficients.
	However, it is not a group under matrix multiplication, as the single-qubit Pauli matrices pick up a factor of $i$ on multiplication - $XY=iZ\notin\PP$.
	The closure of the Pauli basis is the Pauli group $\Pi^{\numqubits}=\{\pm i\}\times\PP^{\numqubits}$; this is four times as large, and no longer has the basis properties of $\PP^{\numqubits}$.
	The Pauli basis inherits a form of multiplication from $\Pi^{\numqubits}$ - $P\cdot Q=R\in\PP^{\numqubits}$ if $PQ=e^{i\phi}R\in\Pi^{\numqubits}$, at which point $\PP^{\numqubits}\equiv D_2^{\numqubits}$.
	However, under this multiplication $\PP^{\numqubits}$ becomes a commutative group, which sacrifices key information about its operator structure.
	Based on the second point in the above list, we may make the following useful definition:
	\begin{dfn}
		The \textbf{relative sign} of $P,Q\in\PP^{\numqubits}$, $s_{P,Q}\in\{-1,1\}$, is defined such that $PQ+s_{P,Q}QP=0$. We further define the \textbf{markers} $\delta_{P,Q}=(1+s_{P,Q})/2$, $\bar{\delta}_{P,Q}=(1-s_{P,Q})/2=1-\delta_{P,Q}$.
	\end{dfn}
	This allows us to write the following useful identity:
	\begin{equation}
		e^{i\theta P}Q=Qe^{is_{P,Q} P}.\label{eq:commute_Pauli_Pauliexp}
	\end{equation}
	Unfortunately this does not extend to the commutation of two such exponentials; one has instead by the application of the Baker-Campbell-Hausdorff formula
	\begin{align}
		e^{i\theta P}e^{i\phi Q}&=e^{i\phi e^{i\theta P/2}Q e^{-i\theta P/2}}e^{i\theta P},\\
		&=e^{i\phi\left[\delta_{P,Q} Q + \bar{\delta}_{P,Q}(\cos(\theta)P+\sin(\theta)PQ)\right]}e^{i\theta P}\\
		&=e^{i\phi Q}e^{i\theta[\delta_{P,Q}P+\bar{\delta}_{P,Q}(\cos(\theta)P+\sin(\theta)PQ)]}.
	\end{align}
	and the exponential expression cannot be simplified unless $\theta=n\pi/2$.
	In this special case, $e^{i\pi/2 P}$ is a Clifford operator (being an operator that maps Pauli operators to Pauli operators); this does not define all Clifford operators, but the set $\{e^{i\pi/2 P},P\in\PP^{\numqubits}\}$ does generate the Clifford group.

	\section{Example of compression over symmetries: the unitary coupled cluster ansatz \label{app:UCC}}
	As an example of symmetry-induced compression, let us construct the Trotterized unitary coupled cluster ansatz~\cite{Peruzzo14,Mcclean16} on a fermionic system. This can be done by taking the Pauli-type ansatz of local Majorana operators acting on an equal number of empty and filled orbitals, removing terms that do not respect $\Kk$, and fixing the remainder to respect the fermion parity. We now detail this procedure.
	
	The UCC ansatz takes the form
	\begin{equation}
	U(\vec{\theta}) = e^{T(\vec{\theta})-T^{\dag}(\vec{\theta})},
	\end{equation}
	where the operator $T(\vec{\theta})$ is a sum of $n$-th order cluster operators $T^{(n)}(\vec{\theta})$ between filled states $i$ and empty states $j$ of the non-interacting problem.
	\begin{equation}
	T^{(n)}(\vec{\theta})=\sum_{i_1,\ldots,i_n,j_1,\ldots,j_n}\theta_{i_1,\ldots,i_n}^{j_1,\ldots,j_n}\hspace{0.1cm}\chatdag_{j_1}\ldots\chatdag_{j_n}\chat_{i_1}\ldots\chat_{i_n}.
	\end{equation}
	The choice of $T(\vec{\theta})-T^{\dag}(\vec{\theta})$ is made to respect $\Kk$ (as creation and annihilation operators are real).
	One typically takes only a few $T^{(n)}$ (usually up to $n=2$), and Trotterizes the resulting expression in terms of individual excitations to implement on a quantum computer, in which case it becomes a product ansatz.
	$\chatdag_j$ and $\chat_j$ are the fermionic creation and annihilation operators for the $j$th orbital.
	These are not themselves Pauli operators, but they may be combined to make Majorana operators
	\begin{equation}
	\gamma_j^{(0)}=\chatdag_j+\chat_j, \gamma_j^{(1)}=i(\chatdag_j-\chat_j),
	\end{equation}
	which are elements of $\PP^{\numqubits}$ (up to a possible sign).
	(One can show this immediately upon choosing a mapping from fermions to qubits.)
	The fermionic number operator, $N=\sum_j\chatdag_j\chat_j$, is equivalent to $\Gamma=\sum_j\gamma_j^{(0)}\gamma_j^{(1)}$ (for commutation purposes).
	To form the operator $T^{(1)}-T^{(1)\dag}$, one may take the set of excitations $e^{i\theta_{i,a}^{j,a}\gamma_i^{(a)}\gamma_j^{(a)}}$ for $i\neq j$ (and $a=0,1$), and enforce the symmetry by fixing $\theta_{i,a}^{j,a}=\theta_{i,1-a}^{j,1-a}$.
	(Terms of the form $\gamma_i^0\gamma_j^1$ do not commute with $\Kk$.)
	The second-order cluster operator is slightly more complicated; one must take all terms of the form
	\begin{equation}
	\exp\left(i\theta_{i_1,i_2,a_1,a_2}^{j_1,j_2,b_1,b_2}\gamma_{i_1}^{a_1}\gamma_{i_2}^{a_2}\gamma_{j_1}^{b_1}\gamma_{j_2}^{b_2}\right),
	\end{equation}
	with $i_1\neq i_2$ ($j_1\neq j_2$) operators for empty (filled) states, and $\sum_ia_i+b_i=1\mod 2$ (terms where $\sum_i a_i+b_i = 0\mod 2$ do not commute with $\Kk$).
	Then, to conserve $\Gamma$, one must fix
	\begin{align*}
	&\theta_{i_1,i_2,0,0}^{j_1,j_2,0,1}=\theta_{i_1,i_2,0,0}^{j_1,j_2,1,0}=-\theta_{i_1,i_2,1,0}^{j_1,j_2,0,0}=-\theta_{i_1,i_2,0,1}^{j_1,j_2,0,0}\\
	&=\theta_{i_1,i_2,0,1}^{j_1,j_2,1,1}=\theta_{i_1,i_2,1,0}^{j_1,j_2,1,1}=-\theta_{i_1,i_2,1,1}^{j_1,j_2,0,1}=-\theta_{i_1,i_2,1,1}^{j_1,j_2,1,0}.
	\end{align*}
	(One can confirm that all operators being fixed commute here, as required.)
	This procedure may be continued as needed to obtain higher-order cluster operators.
	
	One might try to use the tools developed above and check if the Trotterized UCC ansatz tightly spans the reduced Hilbert space. 
	On the one hand, the number of parameters in the full UCC,
	\begin{equation}
	\sum_{n=1}^{\eta}\frac{\eta!}{(\eta-n)!n!}\frac{(\numqubits-\eta)!}{(\numqubits-\eta-n)!n!}=\frac{\numqubits!}{(\numqubits-\eta)!\eta!}-1,
	\end{equation}
	does match precisely the dimension of a real Hilbert space with $\eta$ particles in $\numqubits$ orbitals.
	On the other hand, as the Trotterized UCC Jacobian is full-rank at $\vec{\theta}=\vec{0}$, we strongly suspect that it spans this Hilbert state.
	However, we did not find a definitive proof of this. 
	In particular, Trotterized UCC is not a stabilizer ansatz, and we have not found an obvious construction of a stabilizer ansatz from UCC.
	
	\section{Multivariate Dyson series}\label{app:proof_k_determines_s}
	
	To prove the statement of Lemma \ref{lem:Psi_k_expression}, we need to analyze the multi-parameter expansion \eqref{eq:PT_expansion} of the ground state $\ket{E_0}$, as a perturbative solution to the corresponding eigenvalue equation
	\begin{gather}
	(H_0+JV)\ket{E_0}=E_0\ket{E_0}.\label{eq:PT_Schroedinger_eq}
	\end{gather}
	
	It proves to be convenient to first find an unnormalized solution $\ket{\tilde{E}_0}$ whose expansion states $\ket{\tilde{\Psi}_{\vec{k}}}$ (cf. \eqref{eq:PT_tilde_expansion} ) obey a special condition:
	\begin{gather}
		\braket{\tilde{\Psi}_{\vec{0}}|\tilde{\Psi}_{\vec{k}}}=\delta_{\vec{k},\vec{0}}.\label{eq:PT_Dyson_normalisation}
	\end{gather}
	
	The properly normalized ground state $\ket{E_0}$ is then to be obtained as $\ket{E_0}=\mathcal{N}\ket{\tilde{E}_0}$, for $\mathcal{N}=(\braket{\tilde{E}_0|\tilde{E}_0})^{-1/2}$.

	To find $\ket{\tilde{\Psi}_{\vec{k}}}$, one can use the Dyson series-like approach. For this, one rewrites \eqref{eq:PT_Schroedinger_eq} as:
	\begin{gather}
	(E^{(0)}_0-H_0)\ket{\tilde{E}_0}=(JV-\Delta)\ket{\tilde{E}_0},\label{eq:PT_Dyson_direct}
	\end{gather}	
	for $E^{(0)}_0$ being the unperturbed ground state energy, and quantity $\Delta$ defined as follows:
	\begin{gather}
	\Delta\equiv(E_0-E^{(0)}_0)=\bra{\tilde{\Psi}_{\vec{0}}}JV\ket{\tilde{E}_0}.\label{eq:PT_Delta_def}
	\end{gather}
	
	Eq.~\eqref{eq:PT_Dyson_direct} can be rewritten as:
	\begin{gather}
	\ket{\tilde{E}_0}=\ket{\tilde{\Psi}_{\vec{0}}}+(E^{(0)}_0-H_0)^{-1} (JV-\Delta)\ket{\tilde{E}_0},\label{eq:PT_Dyson_inverse}
	\end{gather}
	
	where the action of the inverse operator $(E^{(0)}_0-H_0)^{-1}$ is well-defined since the state $(JV-\Delta)\ket{\tilde{E}_0}$ has no overlap with $\ket{\tilde{\Psi}_{\vec{0}}}$ (cf. \eqref{eq:PT_Delta_def} and \eqref{eq:PT_Dyson_normalisation}).
	Using expansion \eqref{eq:PT_tilde_expansion} and the form of perturbation $JV=\vec{J}\cdot\vec{V}$, one recovers from \eqref{eq:PT_Dyson_inverse} a set of equations on $\ket{\tilde{\Psi}_{\vec{k}}}$ for all $\vec{k}\neq\vec{0}$:
	\begin{gather}
	\ket{\tilde{\Psi}_{\vec{k}}}=G_0 \left(\sum_{\beta}V_\beta\ket{\tilde{\Psi}_{\vec{k}-\vec{\delta}_{\beta}}}-\sum_{\vec{k}'+\vec{k}''=\vec{k}}\Delta_{\vec{k}'}\ket{\tilde{\Psi}_{\vec{k}''}}\right),\label{eq:PT_iterative_multiJ}\\
	G_0\equiv(E^{(0)}_0-H_0)^{-1},\mathrm{   }\Delta_{\vec{k}}\equiv\sum_{\beta}\bra{\tilde{\Psi}_{\vec{0}}}V_\beta\ket{\tilde{\Psi}_{\vec{k}-\vec{\delta}_\beta}},\label{eq:PT_Delta_multi_definition}
	\end{gather}
	
	for $\vec{\delta}_{\beta}$ the unit vector with the $\beta$ component equal to 1. Note, that the action of $G_0$ here is again well-defined, since it acts on a state which has a zero overlap with $\ket{\tilde{\Psi}_{\vec{0}}}$ (cf. \eqref{eq:PT_Delta_multi_definition} and \eqref{eq:PT_Dyson_normalisation}). Now, with \eqref{eq:PT_iterative_multiJ}, we expressed each state $\ket{\tilde{\Psi}_{\vec{k}}}$ in terms of states $\ket{\tilde{\Psi}_{\vec{k}'}}$ which belong to lower PT orders: $|\vec{k}'|<|\vec{k}|$. Using \eqref{eq:PT_iterative_multiJ} and the unperturbed ground state $\ket{\tilde{\Psi}_{\vec{0}}}=\ket{\vec{0}}$, one can obtain all the states $\ket{\tilde{\Psi}_{\vec{k}}}$ up to any desired order.
	
	Given the states $\ket{\tilde{\Psi}_{\vec{k}}}$, one can also find the expression for the normalization $\mathcal{N}$, as a multi-parameter series:	
	\begin{gather}
	\mathcal{N}=\sum_{\vec{k}}\mathcal{N}_{\vec{k}}\vec{J}^{\cdot\vec{k}} \label{eq:PT_N_k_expansion}
	\end{gather}	
	The expansion states $\ket{\Psi_{\vec{k}}}$ of the normalised ground state $\ket{E_0}$ are then given by:	
	\begin{gather}
	\ket{\Psi_{\vec{k}}}=\sum_{\vec{k}'+\vec{k}''=\vec{k}}\mathcal{N}_{\vec{k}''}\ket{\tilde{\Psi}_{\vec{k}'}} \label{eq:normalising_Psi_k}
	\end{gather}
	
	With this scheme for finding the expansion states $\ket{\Psi_{\vec{k}}}$, we're ready to prove Lemma \ref{lem:Psi_k_expression}. To do so, first we will use \eqref{eq:PT_iterative_multiJ} and prove the validity of the expression \eqref{eq:Psi_tilde_k_expression}, together with the recursive relation \eqref{eq:tilde_Ck}. Then, using \eqref{eq:normalising_Psi_k}, we will extend our proof also to the states $\ket{\Psi_{\vec{k}}}$, recovering the statement of Lemma \ref{lem:Psi_k_expression}.

	\emph{Proof ---} We start with a proof of the relation \eqref{eq:Psi_tilde_k_expression} for the states $\ket{\tilde{\Psi}_{\vec{k}}}$, by induction in PT order $|\vec{k}|$.
	We first note that for $|\vec{k}|=0$, we have a single state $\ket{\tilde{\Psi}_{\vec{k}=\vec{0}}}=\ket{\vec{0}}$ that clearly satisfies \eqref{eq:Psi_tilde_k_expression} - this will be the base of our induction.
	Next, we have to prove \eqref{eq:Psi_tilde_k_expression} for $\ket{\tilde{\Psi}_{\vec{k}}}$ with an arbitrary $\vec{k}$, assuming the validity of \eqref{eq:Psi_tilde_k_expression} for all $\ket{\tilde{\Psi}_{\vec{k}'}}$ s.t. $|\vec{k}'|<|\vec{k}|$. To do so, let us express $\ket{\tilde{\Psi}_{\vec{k}}}$ using \eqref{eq:PT_iterative_multiJ} and show that the different terms that are present on the r.h.s. are proportional to the state $\vec{V}^{\cdot{\vec{k}}}\ket{\vec{0}}$ with a real coefficient. The terms of the type $G_0V_\beta\ket{\tilde{\Psi}_{\vec{k}-\vec{\delta}_{\beta}}}$, assuming expression \eqref{eq:Psi_tilde_k_expression} for $\ket{\tilde{\Psi}_{\vec{k}-\vec{\delta}_{\beta}}}$, can be rewritten as:
	\begin{align}
	G_0V_\beta\ket{\tilde{\Psi}_{\vec{k}-\vec{\delta}_{\beta}}}&= G_0\tilde{C}_{\vec{k}-\vec{\delta}_{\beta}} V_\beta\vec{V}^{\cdot(\vec{k}-\vec{\delta}_{\beta})}\ket{\vec{0}}\\
	&=\frac{S_{\vec{\delta}_\beta,\vec{k}-\delta_\beta} \tilde{C}_{\vec{k}-\vec{\delta}_{\beta}}}{E^{(0)}_{\vec{0}}-E^{(0)}_{\vec{s}(\vec{k})}} \vec{V}^{\cdot{\vec{k}}}\ket{\vec{0}}.\label{eq:PT_induction_term1}
	\end{align}
	The other contributions to the r.h.s. of \eqref{eq:PT_iterative_multiJ} are of the form $G_0\Delta^{(\vec{k}')}\ket{\tilde{\Psi}_{\vec{k}''}}$, such that $\vec{k}'+\vec{k}''=\vec{k}$. The factor $\Delta^{(\vec{k}')}$ here can be rewritten using the assumption of induction:
	\begin{align}
	\Delta_{\vec{k}'}&=\sum_{\beta}\bra{\vec{0}}\tilde{C}^{(\vec{k}'-\vec{\delta}_{\beta})} V_\beta\vec{V}^{\cdot(\vec{k}'-\vec{\delta}_{\beta})}\ket{\vec{0}}\\&=\left(\sum_\beta S_{\vec{\delta}_\beta,\vec{k}-\delta_\beta}\tilde{C}_{\vec{k}'-\vec{\delta}_{\beta}}\right)\bra{\vec{0}}\vec{V}^{\cdot\vec{k}'}\ket{\vec{0}}
	\\&=\Delta_{\vec{k}'}^{\mathrm{Re}}\bra{\vec{0}}\vec{V}^{\cdot\vec{k}'}\ket{\vec{0}},
	\end{align}
	where we introduced the shorthand notation $\Delta_{\vec{k}'}^{\mathrm{Re}}$ for the real coefficient $\left(\sum_\beta S_{\vec{\delta}_\beta,\vec{k}-\delta_\beta}\tilde{C}^{(\vec{k}'-\vec{\delta}_{\beta})}\right)$. With this observation about $\Delta_{\vec{k}'}$ and the assumption of induction at hand, the following manipulation can be performed:
	\begin{align}
	G_0\Delta_{\vec{k}'}&\ket{\tilde{\Psi}_{\vec{k}''}}=\Delta_{\vec{k}'}^{\mathrm{Re}}\tilde{C}_{\vec{k}''}G_0\vec{V}^{\cdot\vec{k}''}\ket{\vec{0}}\bra{\vec{0}}\vec{V}^{\cdot\vec{k}'}\ket{\vec{0}}
	\\&=\frac{\Delta_{\vec{k}'}^{\mathrm{Re}}\tilde{C}_{\vec{k}''}S_{\vec{k}'',\vec{k}'}}{E^{(0)}_{\vec{0}}-E^{(0)}_{\vec{s}(\vec{k})}}\delta_{\vec{s}(\vec{k}'),\vec{0}}\vec{V}^{\cdot\vec{k}}\ket{\vec{0}},\label{eq:PT_induction_term2}
	\end{align}
	where we used the condition $\vec{k}'+\vec{k}''=\vec{k}$. Combining \eqref{eq:PT_induction_term1} and \eqref{eq:PT_induction_term2}, we see that the expression \eqref{eq:PT_iterative_multiJ} indeed implies the form \eqref{eq:Psi_tilde_k_expression} of $\ket{\tilde{\Psi}_{\vec{k}}}$, with a real coefficient $\tilde{C}_{\vec{k}}$ which is given by the formula \eqref{eq:tilde_Ck}.
	
	Before extending this result to the coefficient states $\ket{\Psi_{\vec{k}}}$ of the normalized ground state $\ket{E_0}=\mathcal{N}\ket{\tilde{E}_0}$, we will need to make an aside and prove the following property of the coefficients $\mathcal{N}_{\vec{k}}$:
	\begin{equation}
	\mathcal{N}_{\vec{k}}=\mathcal{N}^{\mathrm{Re}}_{\vec{k}}\bra{\vec{0}}\vec{V}^{\cdot\vec{k}}\ket{\vec{0}},\label{eq:PT_N_k_vacuum_contribution}
	\end{equation}
	for a real coefficient $\mathcal{N}^{\mathrm{Re}}_{\vec{k}}$. First, one can observe that an analogous property holds for the coefficients $Z_{\vec{k}}$ of $Z\equiv\braket{\tilde{E}_0|\tilde{E}_0}=\mathcal{N}^{-2}$:
	\begin{align}
	Z&=\sum_{\vec{k}}\vec{J}^{\cdot{\vec{k}}}Z_{\vec{k}},\\	
	Z_{\vec{k}}&=\sum_{\vec{k}'+\vec{k''}=\vec{k}}\tilde{C}_{\vec{k}'}\tilde{C}_{\vec{k}''}
	\bra{\vec{0}}\left(\vec{V}^{\cdot\vec{k}''}\right)^\dag \vec{V}^{\cdot\vec{k}'}\ket{\vec{0}}
	\\&=\sum_{\vec{k}'+\vec{k''}=\vec{k}}i^{2\Gamma_{\vec{k}''}}S_{\vec{k}'',\vec{k}'}\tilde{C}_{\vec{k}'}\tilde{C}_{\vec{k}''}\bra{\vec{0}}\vec{V}^{\cdot\vec{k}}\ket{\vec{0}}
	\\&=Z^{\mathrm{Re}}_{\vec{k}}\bra{\vec{0}}\vec{V}^{\cdot\vec{k}}\ket{0},\label{eq:PT_Z_k_vacuum_contribution}
	\end{align}
	with a real coefficient $Z^{\mathrm{Re}}_{\vec{k}}$ defined as $\sum_{\vec{k}'+\vec{k''}=\vec{k}}i^{2\Gamma_{\vec{k}''}}S_{\vec{k}'',\vec{k}'}\tilde{C}_{\vec{k}'}\tilde{C}_{\vec{k}''}$; in this derivation, we used \eqref{eq:Psi_k_expression} for states $\ket{\tilde{\Psi}_{\vec{k}}}$. Now, observe that $Z_{\vec{0}}=1$, which means that the norm $\mathcal{N}=Z^{-1/2}=\left(1+\epsilon\right)^{-1/2}$ can be expressed as a Taylor series in $\epsilon=\sum_{\vec{k}\neq0}\vec{J}^{\cdot{\vec{k}}}Z_{\vec{k}}$, which is a quantity of order $O(J)$. Expanding the terms of such Taylor series, one observes that the coefficients $\mathcal{N}_{\vec{k}}$ are given in terms of products of coefficients $Z_{\vec{k}}$ such that the combined perturbation theory order $\vec{k}$ is conserved - for example, a product $Z_{\vec{k}_1}Z_{\vec{k}_2}$ will contribute to $\mathcal{N}_{\vec{k}_1+\vec{k}_2}$. This allows to obtain the property \eqref{eq:PT_N_k_vacuum_contribution} from \eqref{eq:PT_Z_k_vacuum_contribution} term by term. For instance, $Z_{\vec{k}_1}Z_{\vec{k}_2}$ is proportional to $\bra{\vec{0}}\vec{V}^{\cdot(\vec{k}_1+\vec{k}_2)}\ket{\vec{0}}$ with a real coefficient:
	\begin{align}
	Z_{\vec{k}_1}Z_{\vec{k}_2}&=Z^{\mathrm{Re}}_{\vec{k}_1}Z^{\mathrm{Re}}_{\vec{k}_2}\bra{\vec{0}}\vec{V}^{\cdot\vec{k}_1}\ket{\vec{0}}\bra{\vec{0}}\vec{V}^{\cdot\vec{k}_2}\ket{\vec{0}}
	\\&=\delta_{\vec{s}(\vec{k_1}),\vec{0}}\delta_{\vec{s}(\vec{k_2}),\vec{0}}S_{\vec{k_1},\vec{k_2}}Z^{\mathrm{Re}}_{\vec{k}_1}Z^{\mathrm{Re}}_{\vec{k}_2}\bra{\vec{0}}\vec{V}^{\cdot(\vec{k}_1+\vec{k}_2)}\ket{\vec{0}}.
	\end{align}
	This statement can be directly extended to any product of multiple $Z_{\vec{k}}$'s, recovering \eqref{eq:PT_N_k_vacuum_contribution}, as desired.
	\par 
	To prove expression \eqref{eq:Psi_k_expression}, we simply use the property \eqref{eq:PT_N_k_vacuum_contribution} and \eqref{eq:Psi_tilde_k_expression} for  $\ket{\tilde{\Psi}_{\vec{k}}}$, in the formula \eqref{eq:normalising_Psi_k}:
	
	\begin{align}
	\ket{\Psi_{\vec{k}}}&=\sum_{\vec{k}'+\vec{k}''=\vec{k}}\mathcal{N}^{(\vec{k}'')}\ket{\tilde{\Psi}_{\vec{k}'}}
	\\&=\sum_{\vec{k}'+\vec{k}''=\vec{k}}\mathcal{N}^{\mathrm{Re}}_{\vec{k}''}\tilde{C}_{\vec{k}'}\vec{V}^{\cdot\vec{k}'}\ket{\vec{0}}\bra{\vec{0}}\vec{V}^{\cdot\vec{k}''}\ket{\vec{0}}
	\\&=\sum_{\vec{k}'+\vec{k}''=\vec{k}}\delta_{\vec{s}(\vec{k}''),\vec{0}}S_{\vec{k'},\vec{k}''}\mathcal{N}^{\mathrm{Re}}_{\vec{k}''}\tilde{C}_{\vec{k}'}\vec{V}^{\cdot\vec{k}}\ket{\vec{0}}
	\\&=C_{\vec{k}}\vec{V}^{\cdot\vec{k}}\ket{\vec{0}},
	\\C_{\vec{k}}&\equiv\sum_{\vec{k}'+\vec{k}''=\vec{k}}\delta_{\vec{s}(\vec{k}''),\vec{0}}S_{\vec{k'},\vec{k}''}\mathcal{N}^{\mathrm{Re}}_{\vec{k}''}\tilde{C}_{\vec{k}'}.
	\end{align}
	This concludes our proof of Lemma \ref{lem:Psi_k_expression}.\qed

	\section{Separability of disconnected contributions}\label{app:lem_disconnected_contributions_proof}
	
	In what follows, we prove Lemma \ref{lem:disconnected_contributions}.
	
	\emph{Proof --- } Consider a disconnected contribution $\ket{\Psi_{\vec{k}}}=C_{\vec{k}}\vec{V}^{\cdot\vec{k}}\ket{\vec{0}}$ to the ground state $\ket{E_0}$ of the Hamiltonian $H=H_0+\vec{J}\cdot\vec{V}$, with a corresponding splitting $\vec{k}=\vec{k}_A+\vec{k}_B$. The two sets of couplings that are activated, respectively, in $\vec{k}_A$ and $\vec{k}_B$, we will denote $A$ and $B$. We also introduce two non-intersecting sets of qubits, $Q_A$ and $Q_B$, such that they include, respectively, the supports of $\vec{k}_A$ and $\vec{k}_B$, and their union $Q_A\cup Q_B$ constitutes the whole set of qubits.

	Let us consider an auxilliary Hamiltonian $H'$, which is equal to $H$ with a constraint $J_i=0$ for all couplings $V_i$ which are not in $A\cup B$. In the PT series for the ground state $\ket{E_0}'$ of such an auxilliary Hamiltonian,
	\begin{align}
	\ket{E_0}'=\sum_{\vec{k}'}\vec{J}^{\cdot\vec{k}'}C'_{\vec{k}'}\vec{V}^{\cdot\vec{k}'}\ket{\vec{0}},\label{eq:disconnected_PT_series}
 	\end{align} 
 	the terms $C'_{\vec{k}'}$ are equal to the corresponding terms $C_{\vec{k}'}$ in the full series \eqref{eq:PT_expansion} -- namely those, where no couplings $V_i$ are activated besides those in $A\cup B$. In particular, \eqref{eq:disconnected_PT_series} still contains the disconnected contribution of interest, $C'_{\vec{k}'=\vec{k}}=C_{\vec{k}}$. 
 	\par On the other hand, $H'$ is a sum of two independent Hamiltonians, defined on subsystems $Q_A$ and $Q_B$:	
	\begin{align}
	H'&=H'_A\otimes\II_{Q_B}+\II_{Q_A}\otimes H'_B,\\
	H'_A&\equiv-\sum_{i\in Q_A}h_iZ_i+\sum_{i\in A}J_iV_i,\\
	H'_B&\equiv-\sum_{i\in Q_B}h_iZ_i+\sum_{i\in B}J_iV_i.
	\end{align} 	
    This implies that the ground state $\ket{E_0}'$, will be a tensor product of the ground states of $H'_A$ and $H'_B$, 
    \begin{align}
    \ket{E_0}'=\ket{E_0}'_{A}\ket{E_0}'_{B}.\label{eq:disconnected_product_GS}
    \end{align}
    
    In turn, the subsystem ground states $\ket{E_0}'_{A}$ and $\ket{E_0}'_{B}$ can themselves be written as PT series in couplings restricted on $A$ and $B$, separately:
    \begin{align}
    \ket{E_0}'_A=\sum_{\vec{k}'_A}\vec{J}^{\cdot\vec{k}'_A}C'_{\vec{k}'_A}\vec{V}^{\cdot\vec{k}'_A}\ket{\vec{0}}_{Q_A},\label{eq:disconnected_PT_series_A}\\
    \ket{E_0}'_B=\sum_{\vec{k}'_B}\vec{J}^{\cdot\vec{k}'_B}C'_{\vec{k}'_B}\vec{V}^{\cdot\vec{k}'_B}\ket{\vec{0}}_{Q_B},\label{eq:disconnected_PT_series_B}
    \end{align}
    whose terms, again, are identical to those in the full series \eqref{eq:PT_expansion}, with only couplings from $A$ ($B$) activated: $C'_{\vec{k}'_A}=C_{\vec{k}'_A}$ ($C'_{\vec{k}'_B}=C_{\vec{k}'_B}$). Combining \eqref{eq:disconnected_PT_series}, \eqref{eq:disconnected_product_GS}, \eqref{eq:disconnected_PT_series_A} and \eqref{eq:disconnected_PT_series_B}, for our term of interest $C_{\vec{k}}$ we obtain the desired relation:
    \begin{align}
    C_{\vec{k}}=C_{\vec{k}_A}C_{\vec{k}_B}.\label{eq:disconnected_contribution_primed}
    \end{align}
    
	\qed
	
	\begin{figure}[H]
	\includegraphics[width=\columnwidth]{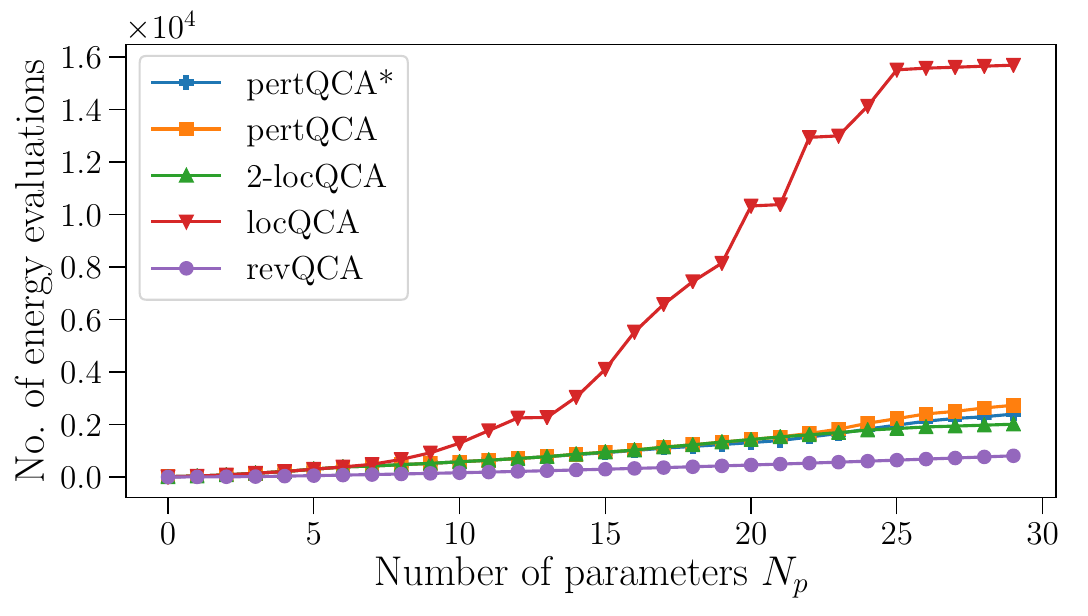}
	\caption{\label{fig:QCA_convergence_weakcoupling}Plot of the optimization convergence speed (Eq.~\eqref{eq:rel_energy_error}) for different variational hierarchies in a weakly-coupled transverse-field Ising model ($J/h=0.15$). Convergence is represented by a total number of energy function evaluations $n_{ev}$ and plotted as a function of the number of parameters used. Note that the optimization of $N_p$ ansatz parameters always uses the optimized value of $N_p-1$ parameters for initialization (see Sec.~\ref{sec:VQE_performance}). Because of this, in $n_{ev}(N_p)$ we always include $n_{ev}(N_p-1)$ and the resulting plots are by definition monotonic.}
	\end{figure}
	
	\begin{figure}[H]
	\includegraphics[width=\columnwidth]{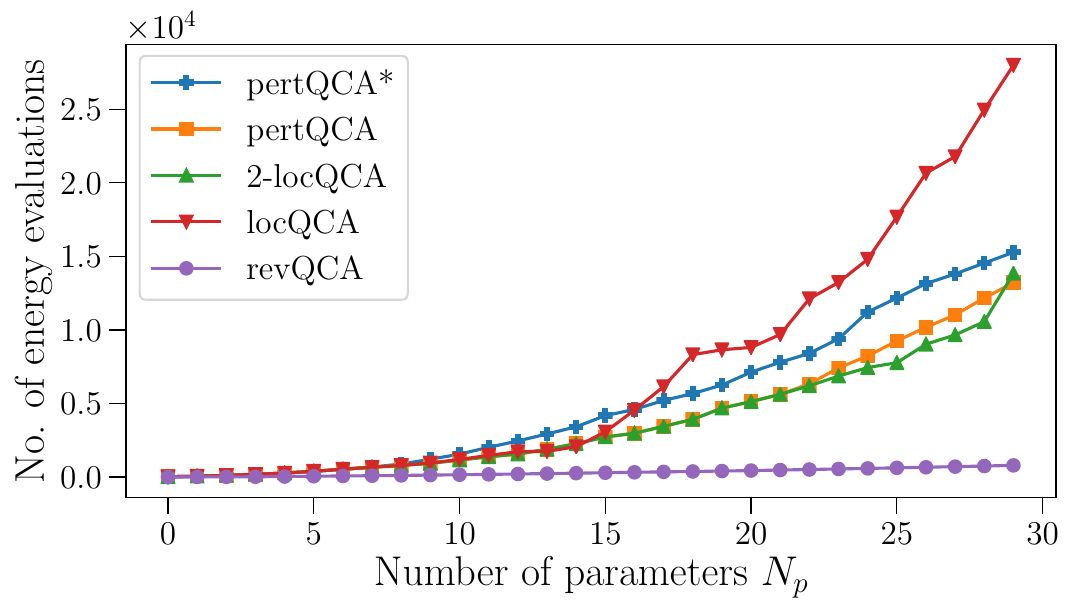}
	\caption{\label{fig:QCA_convergence_strongcoupling} Plot of convergence speed similar to Fig.~\ref{fig:QCA_convergence_weakcoupling}, but in the strongly-coupled regime instead ($J/h=6$).}
	\end{figure}
	
	\begin{figure}[H]
	\includegraphics[width=\columnwidth]{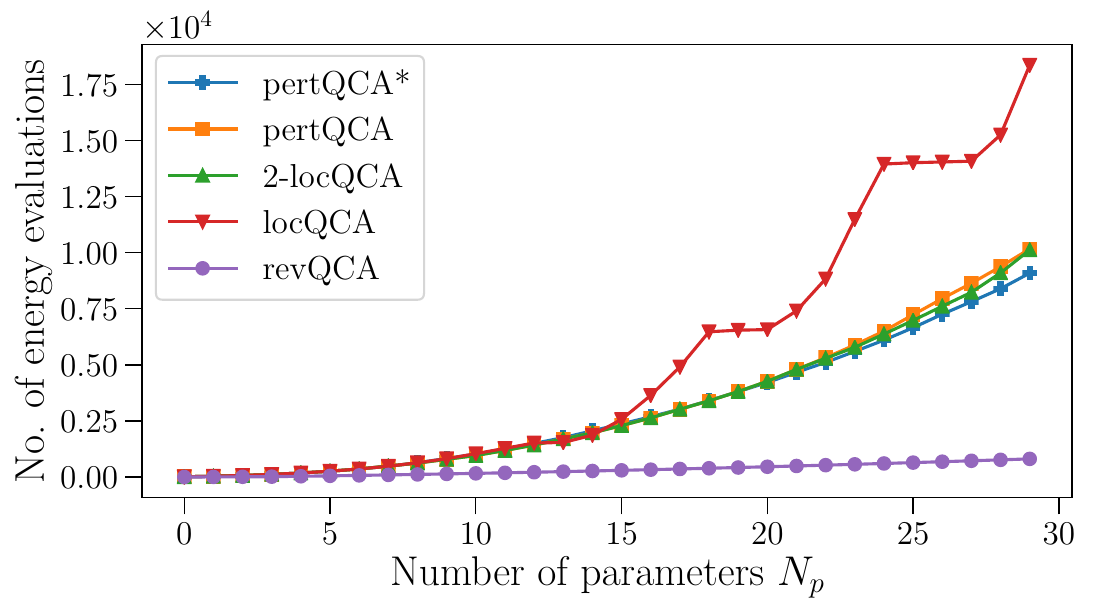}
	\caption{\label{fig:QCA_convergence_critical} Plot of convergence speed similar to Fig.~\ref{fig:QCA_convergence_weakcoupling}, but in the critical regime instead ($J/h=1$).}
	\end{figure}
	
	\section{Convergence speed of classical optimization of QCA}

	In this appendix we show the convergence rate of our classical optimization of QCA in terms of the number of function evaluations for Fig.~\ref{fig:QCA_performance_weakcoupling}, Fig.~\ref{fig:QCA_performance_strongcoupling} and Fig.~\ref{fig:QCA_performance_critical} (Fig.~\ref{fig:QCA_convergence_weakcoupling}, Fig.~\ref{fig:QCA_convergence_strongcoupling} and Fig.~\ref{fig:QCA_convergence_critical} respectively).
	We have not performed any metaparameter tuning for this optimization, which would likely improve these numbers significantly.
	The optimization here was performed in the absence of realistic conditions on quantum hardware (in particular in the absence of sampling noise); any further optimization of convergence times would need to take this into account in order to make a realistic comparison to other ansatzes.

\end{document}